\documentclass[usegraphicx]{mn2e}

\usepackage{graphicx}
\usepackage{afterpage}

\newcommand\Msun{{\,M_\odot}}

\title[Galaxy Merger Morphologies and Time-Scales]{Galaxy Merger Morphologies and Time-Scales
 \\ from Simulations of Equal-Mass Gas-Rich Disc Mergers}

\author[J.M. Lotz et al.]{Jennifer M. Lotz,$^1$\thanks{NOAO Leo Goldberg Fellow},  
Patrik Jonsson,$^2$ T.J. Cox,$^3$\thanks{W.M. Keck Fellow} and Joel R. Primack$^2$ \\
$^1$ National Optical Astronomical Observatory, 950 N.  Cherry Avenue, Tucson, AZ 85719, USA; lotz@noao.edu \\
$^2$ Department of Physics, University of California, Santa Cruz, CA, USA, 95064 \\
$^3$ Harvard-Smithsonian Center for Astrophysics, 60 Garden St.,  Cambridge, MA, 02138 USA}

\begin{document}

\date{Submitted 2008 8 May}

\pagerange{\pageref{firstpage}--\pageref{lastpage}} \pubyear{2008}

\maketitle

\label{firstpage}

\begin{abstract}
A key obstacle to understanding the galaxy merger rate and its role
in galaxy evolution is the difficulty in constraining the merger 
properties and time-scales from instantaneous snapshots of the real universe.
The most common way to identify galaxy mergers is by morphology, yet current
theoretical calculations of the time-scales for galaxy disturbances are quite
crude.  We present a morphological analysis of a large suite of GADGET N-Body/hydro-dynamical 
equal-mass gas-rich disc galaxy mergers which have been processed through
the Monte-Carlo radiative transfer code SUNRISE.  With the resulting images,
we examine the dependence of quantitative morphology ($G$, $M_{20}$, $C$, $A$) 
in the SDSS $g$-band on  merger stage, dust, viewing angle, orbital parameters, gas properties, 
supernova feedback, and total mass.   We find that mergers appear most disturbed in
$G-M_{20}$ and asymmetry at the first pass and at the final coalescence of their nuclei, 
but can have normal quantitative morphologies at other merger stages.   The merger
observability time-scales depend on the method used to identify the merger as well
as the gas fraction, pericentric distance, and relative orientation of the merging
galaxies.  Enhanced star formation peaks after and lasts significantly longer than 
strong morphological disturbances.   Despite their massive bulges,  the 
majority of merger remnants appear disc-like and dusty in $g$-band light 
because of the presence of a low-mass star-forming disc. 
\end{abstract}


\section{INTRODUCTION}
It remains unknown to what degree present-day galaxies are assembled discretely via
the merger of pre-existing galaxies (e.g. Kauffmann, White, \& Guideroni 1993; Somerville,
Primack, \& Faber 2001) or through more continuous processes such as cold gas 
and dark matter accretion (e.g. Kere\v{s} et al. 2005; Dekel \& Birnboim 2006).  
The most obvious way to constrain the importance of galaxy mergers
is to count the number of on-going merger events. However, current observational constraints on 
the galaxy merger rate are highly  uncertain and strongly debated (Ryan et al. 2008; Lin et al. 2008; 
Lotz et al. 2008;  Renzini 2007;  Kartaltepe et al. 2007;  Masjedi et al. 2008).  
Moreover, theoretical predictions for the galaxy merger rate and mass assembly can vary by 
factors of ten (Jogee et al. 2008).   These discrepancies are partially the result of the non-trivial
conversion of the observed number density of galaxy mergers into a galaxy merger rate and 
the comparison of this galaxy merger rate to the cosmological predictions for dark matter
halo assembly (e.g. Berrier et al. 2007; Guo \& White 2008; Kitzbichler \& White 2008). 

Since the realization that the merger of two disc galaxies
could produce a spheroidal galaxy (Toomre 1977), spheroidal galaxies, red galaxies, 
and post-starburst galaxies have been used
to indirectly trace the role of galaxy mergers in galaxy evolution (e.g. Bell et al. 2004; 
Faber et al. 2007; Brown et al. 2007; Abraham et al. 2007; Hopkins et al. 2007; Hogg et al. 2006).
However, there are multiple ways to quench star-formation and produce post-starbursts 
or red spheroidal galaxies (e.g.  Dekel \& Birnboim 2006; Moore, Lake, \& Katz et al. 1998). 
Also, simulations of galaxy mergers often show significant star-formation and disc components
well after the merger event (Springel \& Hernquist 2005; Robertson et al. 2006a; 
Naab, Jesseit, \& Burkert 2006; Cox et al. 2006, 2008; Khalatyan et al 2008), thus the 
photometric and morphological signatures of merger remnants
are ambiguous. A merger remnant's star-formation history, morphology, and
kinematics are likely to depend on the properties of the progenitors
and the merger conditions,  which are increasingly difficult to determine
as time passes. Therefore {\it direct} observations of on-going 
galaxy mergers are needed to constrain the role of mergers in galaxy assembly.  

Galaxy merging is a process that lasts several billion years rather than a short-lived event.  
The signatures of a galaxy merger change with merger stage, making the
identification of galaxy mergers challenging.  If hierarchical models of galaxy assembly are correct, 
then the majority of massive galaxies could be considered an on-going merger or a merger
remnant which will undergo another
merger event within a few Gyr (Stewart et al. 2008).  We shall define a galaxy merger 
as a pair of galaxies which 
are gravitationally bound and whose orbits will dynamically decay such that their nuclei 
will merge within $x$ billion years, where $x$ is typically 1-3 Gyr for `major' mergers with mass
ratios greater than 1:3. 

There are two general approaches to identifying such systems observationally. 
The first approach is to find close pairs of galaxies before their nuclei have coalesced, 
either by selecting galaxies close in projected angular separation and line-of-sight radial 
velocity (e.g. Patton et al. 2000; Barton, Geller, \& Kenyon 2000; Lin et al. 2004; 
de Propris et al. 2005; Lin et al. 2008) or by measuring the deprojected correlation function 
on small scales (Bell et al. 2006a; Masjedi et al. 2006, 2008; 
Li et al. 2008).  The second approach is
to identify morphologically disturbed galaxies, some of which will be post-mergers and
some of which will be interacting pairs.   Morphological disturbances can be found qualitatively
through visual inspection (e.g. Kampczyk et al. 2007; Bundy et al. 2005;  Brinchmann et al. 1998), 
or by quantitative measures such as the Gini coefficient, second-order
moment of the brightest 20\% of the light ($M_{20}$), and asymmetry (Abraham et al. 1994; Conselice 2003;
Abraham et al. 2003; Lotz, Primack, \& Madau 2004 [LPM04];  Scarlata et al. 2007).   
At present, all quantitative merger indicators are calibrated empirically using galaxies 
with `normal' and `disturbed' visual classifications.

Translating the number of observed merger candidates into a merger rate requires the assumption
of an observability time-scale -- the time during which one would have identified the
system as merging.  Until now, this time-scale has been poorly constrained. 
Close pair studies often assume a dynamical friction time-scale (which varies from 200 Myr to 1 Gyr).   
However this value does not take into account the range of possible orbits for the merging 
system nor the time during which the system would not meet the pair criteria at very large 
and very small separations. Also, recent comparison of N-body simulations to analytical calculations
indicate that analytically-derived dynamical friction time-scales can deviate substantially from
those predicted by N-body simulations (Boylan-Kolchin, Ma, \& Quataert 2008; Jiang et al. 2008).  
Very similar time-scales are also generally assumed for mergers selected using
both visual and quantitative morphologies.  This assumption is even less likely to be valid 
given that different morphological selection criteria are sensitive to different stages of 
the merger process.  For example, visual classification using a combination of signatures
 (e.g. tidal tails, multiple nuclei, shells) is likely to be sensitive for longer time-scales 
and lower-mass merger ratios than current quantitative methods. 

The observability time-scale for a particular merger may depend on (1) the method used to identify the
merger; (2) the merger parameters (mass ratio, gas properties, bulge/disc ratio, orbits, dust content);
and (3) the observational selection (observed wavelength, viewing angle, spatial resolution).
Cosmological-scale numerical simulations currently do not have the spatial resolution to directly determine the 
cosmologically-averaged observability time-scale for each method.  Therefore, one is required to
 use a suite of 
galaxy-scale numerical simulations which span a large range of input merger parameters to 
constrain the observability time-scales for the different input parameters. 
Given a sufficiently broad range of merger parameters,  the observability time-scales for each parameter set
may then be weighted by the probability distribution of the mass ratios, gas fraction, etc. which can be computed
from current cosmological-scale simulations.  An additional complication is that 
galaxy-scale numerical simulations typically 
track the distribution of `particles' (star, gas, and dark matter), as opposed to 
the projected light distribution at a particular wavelength (which is what is observed). 
A few works have attempted to quantify the observability time-scales using only the stellar particles
(Conselice 2006; Bell et al. 2006b) or gas particles (Iono, Yun, \& Mihos 2004) 
and ignoring the effects of dust and age-dependent stellar luminosities.  While
this may be acceptable for gas-poor dissipationless mergers (e.g. Bell et al. 2006b), the 
appearance of most gas-rich mergers is almost certainly affected strongly by both young stars and dust.

In this work, we present a first attempt to constrain the observability time-scales for a variety of
methods for identifying galaxy mergers.  We present a morphological analysis of a large suite of 
GADGET N-Body/SPH
equal-mass gas-rich disc galaxy merger simulations which have been processed through
the Monte-Carlo radiative transfer code SUNRISE.  With the resulting images,
we examine the dependence of quantitative morphology in the SDSS $g$-band ($\lambda_c =$ 4686\AA) 
on  merger stage, dust, viewing angle, orbital parameters, gas properties, supernova
feedback, and total mass.    We constrain the time-scales of quantitative
morphology disturbances in Gini coefficient, $M_{20}$, and asymmetry, 
and the time-scales during which close pairs lie at projected separations 
$R_{proj} < $20, 30, 50, and 100 $h^{-1}$ kpc, assuming $h = 0.7$. 
Finally, we compare the simulated merger remnant morphologies and star-formation rates. 
In \S2, we describe the galaxy merger simulations, including the GADGET N-Body/SPH calculations,
the SUNRISE Monte-Carlo radiative transfer calculations, the initial galaxy models, and the
range of merger parameters explored.  In \S3, we define the morphological quantities 
$G$, $M_{20}$, $C$  and $A$.
We also describe the different observational criteria used to identify galaxy mergers, and define
the merger observability time-scale for each method.  In \S4, we present the results of our analysis, and
in \S5 we discuss the implications of these results for finding galaxy mergers, calculating the
merger rate, and the properties of merger remnants.
A subsequent paper with a similar analysis of non-equal mass mergers is in preparation.

\section{GALAXY MERGER SIMULATIONS}
\subsection{GADGET N-Body/SPH simulations}
All of the simulations presented in this work were performed using the 
N-Body/SPH code GADGET (Springel et al. 2001).  The details of these simulations, 
their global star-formation histories, and their remnant properties are discussed
in Cox et. al (2004, 2006, 2008). Each galaxy is initially modeled as a disc of
stars and gas, a stellar bulge, and a dark matter halo. The stellar and dark matter particles
are collisionless and are subject to only gravitational forces.  The gas particles are also
subject to hydro-dynamical forces. The baryonic and dark matter particles have gravitational
softening lengths of 100 pc and 400 pc respectively.  The SPH smoothing length for the
gas particles indicates the size of the region over which the particle's hydrodynamic quantities are
averaged and is required to be greater than half the gravitational softening length or
$>$ 50 pc.   While we use the first version of GADGET (Springel et al. 2001), the smoothed
particle hydrodynamics are upgraded to use the `conservative entropy' version that
is described in Springel \& Hernquist (2002). The radiative cooling rate $\Lambda_{net}$($\rho$, $u$)
is computed for a primordial plasma as described in Katz et al. (1996).  

\begin{figure*} 
  \includegraphics[width=160mm]{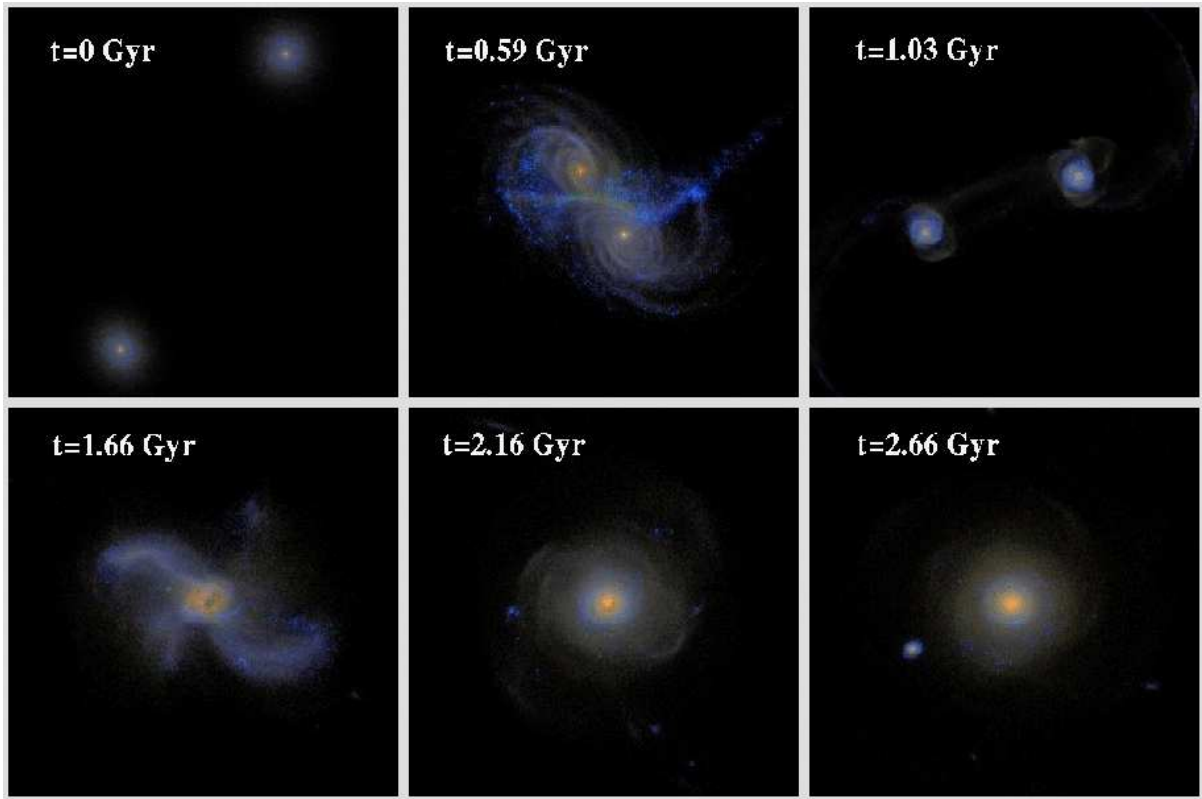}
  \caption{SDSS $u-r-z$ composite colour images with dust extinction 
for the high-resolution Sbc prograde-prograde simulation (SbcPPx10) as viewed by camera 0 (face-on). Time since the
start of the simulation in given in the upper left corner of each image. Shown in
the top row are the initial pre-merger galaxies, the first
pass, the maximal separation after the first pass, and in the bottom row are the merger of the nuclei, the post-merger at
0.5 Gyr after the merger, and the remnant at 1 Gyr after the merger. The field of view for the initial galaxies
and the maximal separation is 200 kpc, while the field of view for the other images is 100 kpc. The merger morphologies are most disturbed at the first pass and merger.  Star-forming regions in the initial discs, tidal tails, and outer regions of the remnant appear blue, while the dust-enshrouded star-forming nuclei 
appear red.}
\end{figure*}

\begin{figure*} 
  \includegraphics[width=160mm]{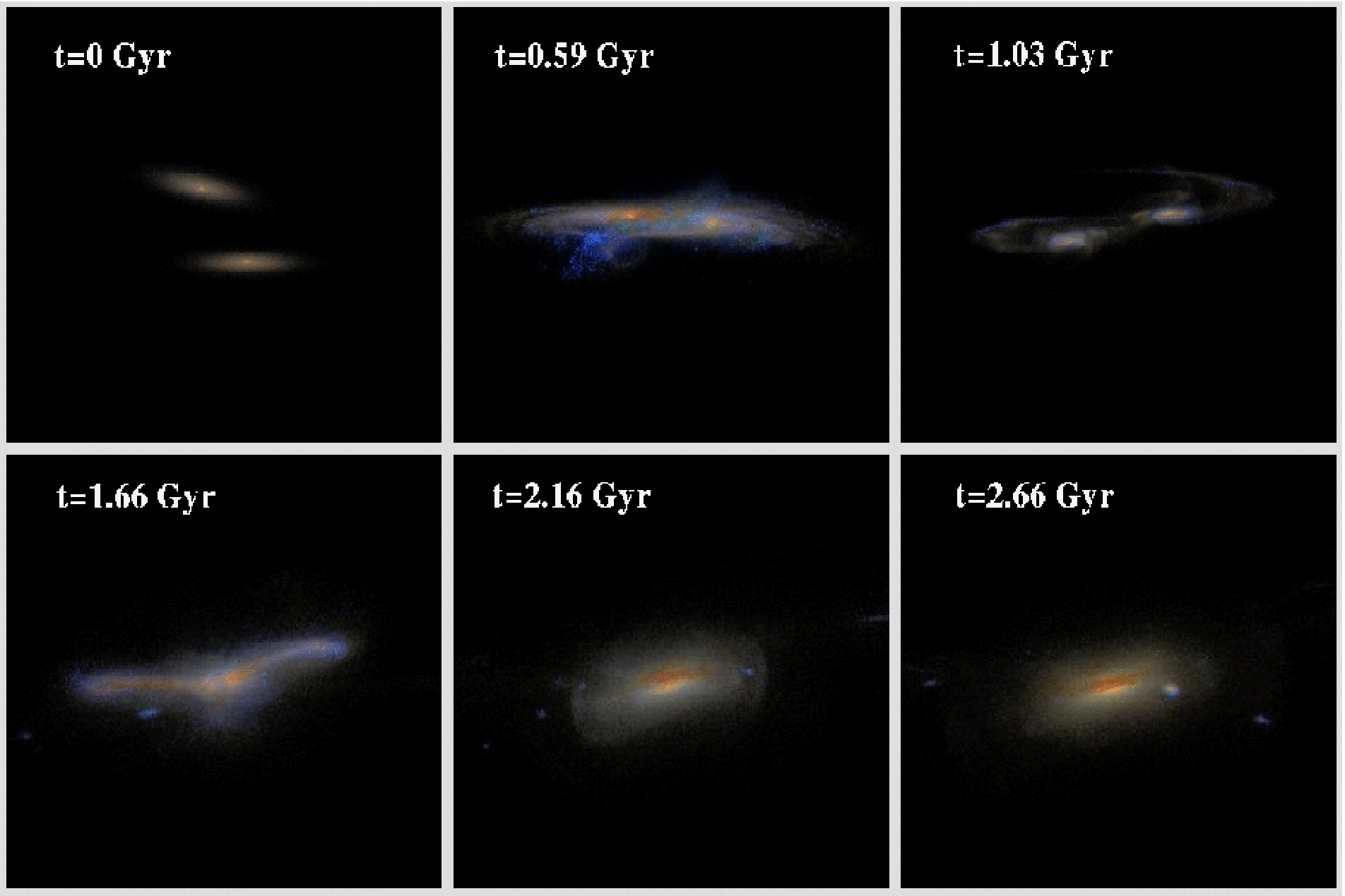}
    \caption{SDSS $u-r-z$ composite colour images for the same simulation as Fig. 1 (SbcPPx10)  as viewed by camera 4 (roughly edge-on). Time since the
start of the simulation in given in the upper left corner of each image. 
The timesteps and image scales are same as the previous figure. When viewed edge-on, the dust lanes associated with initial discs and remnants are clearly visible.}
\end{figure*}

Gas particles are transformed into collisionless star particles assuming the
Kennicutt-Schmidt law (Kennicutt 1998) where the star-formation rate depends
on the local gas density $\rho_{gas}$.  This occurs in a stochastic sense
(Springel \& Hernquist 2003) in which each gas particle can spawn one or two new star
particles with a probability determined by the star-formation rate.  
These new star particles have typical masses $\sim 10^5$ M$_{\odot}$, and are assigned
ages based on their formation time and metallicities based on the metallicity of
the gas particle from which they are spawned.  We adopt the instantaneous recycling approximation
for metal production whereby massive stars are assumed to instantly become supernovae, 
and the metals produced are put back into the gas phase of the particle. 
In this version of GADGET, metals do not mix and remain in the gas particle in which they
are formed. The enriched gas contribution from stellar winds and Type Ia supernovae are ignored. 
Unlike the metals, there is no recycling of hydrogen and helium to the gas.  

Feedback from supernovae is required to produce stable star-forming discs. 
Energy from supernovae heats and pressurizes the interstellar gas and stabilizes it
against gravitational collapse.  Because of the limited resolution of most N-Body/SPH
simulations, the physical processes associated with supernova feedback cannot
be directly modeled and must be included using simple prescriptions.  We test two
supernova feedback models, which are discussed in detail in Cox et al. (2006).  
Both models assume the supernova feedback energy is
dissipated on an 8 Myr time-scale, and have a equation of state parameterized by
$n$, where $P \sim \rho_{gas}^{1 + (n/2)}$.  The $n=2$ model treats star-forming gas with
a stiff equation of state where the pressure in star-forming regions scales as
$P \sim \rho^2_{gas}$, while the $n=0$ model assumes that this
gas is isothermal with an effective temperature $\sim 10^5$ K ($P \sim \rho_{gas}$).   
Both feedback models
produce stable isolated star-forming discs and predict similar gas consumption during the merger, 
but the strength and timing of the merger-induced starbursts depend on the feedback model assumed 
(Cox et al. 2006). No active galactic nuclei are included in these simulations.  Such AGN are expected
to influence the star-formation and morphologies only during the post-merger stages (see \S 5
for discussion). 

\subsection{SUNRISE Monte Carlo radiative transfer processing}
SUNRISE is a parallel code which performs full Monte Carlo radiative-transfer calculations
using an adaptive-mesh refinement grid (Jonsson 2006; Jonsson et al. 2006).  This code was developed 
to calculate
the effects of dust on the emission from the GADGET N-Body/SPH simulations. 
SUNRISE can model arbitrary geometries of emitting and absorbing/scattering material
with a large spatial dynamical range and efficiently generate images of the emerging 
radiation viewed from arbitrary points in space.  We use SUNRISE v2 for these simulations (Jonsson 2006).
Given a particular simulation geometry and viewing angle, SUNRISE v2 
performs the Monte-Carlo radiative transfer calculation for
20 wavelengths from the far-ultraviolet to the mid-infrared and interpolates a resulting spectral
energy distribution of 510 wavelengths including the effects of absorption and scattering.  
 
At least 30 timesteps are analysed for each merger simulation.  
For each GADGET simulation timestep,  SUNRISE assigns a spectral energy distribution
to each star particle using the STARBURST99 population synthesis models (Leitherer et al. 1999).  
New star particles are treated as single stellar populations with ages based on their formation time. 
Bulge star particles form in an instantaneous burst well before the start of the simulation 
(8-13 Gyr depending on the galaxy model; see Rocha et al. 2008). Initial disc star particles 
are assumed to have formed with an 
exponentially declining star-formation starting at the time of the formation of the bulge. 
The metallicities of the gas and stars of the initial galaxy models decline
exponentially with the radius of the disc.  The density of dust is linearly proportional
to the density of metals in the gas.  

The dust attenuation in the initial galaxy models have been found to reproduce the observed galaxy 
inclination-attenuation relations and the global infrared-to-ultraviolet flux ratios for spiral 
galaxies (Rocha et al. 2008).   During the merger,  the gas discs of the initial galaxies become
disrupted, resulting in complicated dust geometries.   The full radiative transfer calculation done by
the SUNRISE v2 code are well-suited to determining the effects of complicated dust geometries.  
However, the dust approximation is limited by the low spatial resolution of the regions of cold gas and star-formation in 
the input GADGET simulations ($\sim$ 100 pc).   Therefore the attenuation of very young stars may
be underestimated,  and the effects of clumpy dust and gas on small scales are not included. A future version of
SUNRISE will include improved treatment of small scales. 
The attenuation and infrared luminosities of the output SUNRISE images during the merger have 
been compared to the available literature for local dusty mergers. 
Jonsson et al. (2006) found that these simulations reproduce the observed relationships 
between ultraviolet spectral slope $\beta$  and the global infrared-to-ultraviolet flux ratio,  
thus the simulated dust distribution is a reasonable approximation of the dust found in local
dusty mergers.

Images in multiple band-passes (GALEX FUV/NUV, SDSS $ugriz$, 2MASS $JHK$) 
for 11 isotropically positioned viewpoints (`cameras') are generated and the
total absorbed bolometric luminosity over all wavelengths/viewing angles is calculated.
These cameras are positioned with respect to the plane of the merger orbit at $\phi$, $\theta$ (degrees) 
= (0,0), (0, 79), (72, 79), (144, 79), (216,79), (288,79), (0, 127), (72, 127), (144, 127), (216, 127), and (288,127).   
In Figures 1 and 2, we show examples of composite SDSS $u-r-z$ images for one of our simulations viewed face-on (camera 0) 
and edge-on (camera 4). 
The predicted dust attenuation for the initial undisturbed galaxy models agrees well with 
observations of dust attenuation in local disc galaxies (Rocha et al. 2008).

\subsection{Initial Galaxy Models}
The goal of this work is to calibrate the morphological disturbance time-scales for merging and
interacting galaxies using realistic `observations' of galaxy merger simulations 
including the effects of star formation and dust.   Dust and star formation have a much stronger effect on
gas-rich merger morphologies than dissipationless mergers, hence we examine the mergers
of gas-rich disc galaxies.  All of the disc galaxy models explored here have relatively small bulge components
with stellar bulge-to-disc mass ratios $\leq$ 0.25.  Such low bulge-to-disc ratio galaxies are more likely to experience
strong starbursts (Mihos \& Hernquist 1996,  Cox et al. 2008) and stronger morphological disturbances (Conselice 2006). 
These galaxies may also be more representative of high redshift mergers, as bulge-dominated systems are increasingly rare
at $z>1$ (e.g. Lotz et al. 2008; Ravindranath et al. 2006).  The structure of dissipationless spheroidal galaxy merger remnants 
have been studied by several other authors (e.g. Naab, Khochfar,\& Burkert 2006;  Boylan-Kolchan, Ma, \& Quartaert 2005).  

We adopt two general models for gas-rich discs: the `Sbc' model tuned to match a large, gas-rich Sbc
disc galaxy, and the `G-series' of discs with varying mass (G3, G3, G1, G0) with lower gas fractions tuned to match
SDSS observations of local galaxies.   (We use the notation of `Sbc' and `G' for these different galaxy models in keeping with
previous publications based on these simulations: Cox et al. 2006, 2008, Jonsson et al. 2006, Rocha et al. 2008). 
Each galaxy model contains a rotationally supported disc of gas and stars, a non-rotating stellar bulge, and
a massive dark matter halo (Table 1). A detailed description of the galaxy disc models can be found in Cox et al. (2006, 2008), 
Jonsson et al. 2006 and Rocha et al. (2008).
\begin{table*}
  \centering
  \begin{minipage}{168mm}
    \caption{Initial Galaxy Conditions}
    \begin{tabular}{@{}lccrllllccccr@{}} 
      \hline
       Model  &  $N_{part}$\footnote{Total number of particles in GADGET simulation for fiducial resolution models.} & $M_{vir}$\footnote{Virial mass} & $C$\footnote{Dark matter halo concentration}  & $M_{bary}$\footnote{Baryonic mass} & $M^*_{disc}$\footnote{Mass of stellar disc} & $M^*_{bulge}$\footnote{Mass of stellar bulge} & $M_{gas}$\footnote{Mass of gaseous disc} & $f_{bulge}$\footnote{Fraction of baryons in the bulge} & $f_{gas}$\footnote{Fraction of baryons in gas} & $R_{disc}$\footnote{Scalelength of stellar disc}  & $R_{bulge}$\footnote{Scalelength of bulge}  &$R_{gas}$\footnote{Scalelength of gaseous disc} \\
     &  & ($\Msun$) &  &($\Msun$) &($\Msun$) & ($\Msun$) & ($\Msun$)  &  &   & (kpc)  & (kpc)  & (kpc) \\
     \hline
 Sbc  & $1.7\cdot10^5$  & $8.1\cdot10^{11}$  &11  & $1.0\cdot10^{11}$  &$3.9\cdot10^{10}$  &$9.7\cdot10^9$     &$5.3\cdot10^{10}$ & 0.10 &0.52  &5.50  & 0.45   & 16.50 \\
 G3   & $2.4\cdot10^5$  & $1.2\cdot10^{12}$  &6   & $6.2\cdot10^{10}$  &$4.1\cdot10^{10}$  &$8.9\cdot10^9$   &$1.2\cdot10^{10}$ & 0.14 &0.19  &2.85  & 0.62   &  8.55 \\
 G2   & $1.5\cdot10^5$  & $5.1\cdot10^{11}$  &9   & $2.0\cdot10^{10}$  &$1.4\cdot10^{10}$  &$1.5\cdot10^9$   &$4.8\cdot10^{9}$  & 0.08 &0.24  &1.91  & 0.43   &  5.73 \\
 G1   & $9.5\cdot10^4$  & $2.0\cdot10^{11}$  &12  & $7.0\cdot10^{9}$   &$4.7\cdot10^{9}$   &$3.0\cdot10^8$   &$2.0\cdot10^{9}$  & 0.04 &0.29  &1.48  & 0.33   &  4.44 \\
 G0   & $5.1\cdot10^4$  & $5.1\cdot10^{10}$  &14  & $1.6\cdot10^{9}$   &$9.8\cdot10^{8}$   &$2.0\cdot10^7$   &$6.0\cdot10^{8}$  & 0.01 &0.38  &1.12  & 0.25   &  3.36 \\
\hline
\end{tabular}
\end{minipage}
\end{table*}

The Sbc model parameters are motivated by observations of local gas-rich, disc-dominated Sbc galaxies similar to the Milky Way
(see Table 1; Cox et al. 2006).  
The optical (stellar) disc scalelength, dynamical mass, and gas fraction are taken from Roberts \& Haynes (1994).  
The bulge-to-disc ratio and bulge size are from the observations of de Jong (1996) and the total stellar mass
is derived using the Bell \& de Jong (2001) relations.  We assume that the gas disc is exponential with a scalelength 
three times the stellar disc scalelength (Broeils \& van Woerden 1994).  We adopt an adiabatically contracted
NFW dark matter halo with a concentration of 11. The resulting Sbc model has a viral mass of 
8.12 $\times 10^{11}$ M$_{\odot}$, with a 12.5\% baryonic mass fraction.  Fifty-two percent of the baryons are in gas
(mostly at large radii), and $\sim$ 10\% of the baryons are bulge stars. 

In order to sample the parameter space spanned by many present-day galaxies, we also explored mergers between model
galaxies with masses, bulge-to-disc ratios, and gas fractions motivated by SDSS estimates of typical local galaxies (Table 1; Cox et al. 2008). 
We refer to these model galaxies as the G-series. 
The largest galaxy (G3) is chosen to have a stellar mass $\sim 5 \times 10^{10}$ M$_{\odot}$, and the smaller galaxies
are chosen to have stellar masses $\sim 1.5 \times 10^{10}$ M$_{\odot}$ (G2),  $0.5 \times 10^{10}$ M$_{\odot}$ (G1),
and $0.1 \times 10^{10}$ M$_{\odot}$ (G0), spanning a factor of 50 in stellar mass.  
The stellar half-light radii are from the stellar mass-size relation
of Shen et al. (2003).  The bulge-to-disc ratios are taken from de Jong (1996) and used to determine the
stellar disc and bulge masses and scalelengths.   The gas fractions and masses are determined from the gas mass
- stellar mass scaling relation from Bell et al. (2003).  As for the Sbc model, the gas scalelength is
assumed to be three times the stellar disc scalelength.   We adopt NFW dark matter halo profiles selected
such that the rotation curves lie on the baryonic Tully-Fisher relation (Bell \& de Jong 2001; Bell et al. 2003).
Unlike the Sbc model, these models do not include adiabatic contraction. The total mass-to-light ratio is
assumed to vary with mass such that lower mass galaxy model have higher mass-to-light ratios and the
range in total mass is a factor of 23.   While the total masses of the Sbc and G3 models are similar,  the G3 model
has a lower gas fraction, a smaller disc scalelength, and consequently, much less gas at large radii than the Sbc model
(Table 1). 

\subsection{Galaxy Merger Parameters}

\begin{table}
  \centering
  \begin{minipage}{84mm}
  \caption{Equal-Mass Merger Simulation Parameters}
  \begin{tabular}{@{}lccrrrrrr@{}}
    \hline
    Simulation & $x$\footnote{The numerical resolution of simulation is $x N_{part}$, where $N_{part}$ is given in Table 1.} & $n$\footnote{Supernova feedback polytropic index $n$ where $P \propto \rho^{1 + (n/2)}$ in star-forming regions.} & $\theta_1$\footnote{Initial orientation of galaxy 1 with respect to the plane of the orbit in spherical coordinates, where $\theta = \arctan (\frac{\sqrt{ x^2 + y^2}}{z})$ and $\phi = \arctan (\frac{y}{x}$)} & $\phi_1^c$ & $\theta_2$\footnote{Initial orientation of galaxy 2} & $\phi_2^d$   & $e$\footnote{Eccentricity of the orbit, where a parabolic orbit has $e=1$.} & R$_{peri}$\footnote{Pericentric distance of the initial orbit} \\
     &   &  &   &  &   &  &   & (kpc) \\
     \hline
     \multicolumn{9}{c}{Sbc-Sbc mergers} \\
     \hline
     SbcPPx10  & 10 & 2  & 0  &0      & 30 & 60 & 1.00  & 11   \\
     SbcPPx4  & 4  & 2  & 0  &0      & 30 & 60  & 1.00  & 11   \\
     \hline
     SbcPP     & 1  & 2  & 0  &0      & 30 & 60 & 1.00  & 11   \\
     SbcPR     & 1  & 2  & 180 &0     & 30 & 60 & 1.00  & 11          \\
     SbcRR     & 1  & 2  & 180 &0    & 210 &60 & 1.00  & 11         \\
     SbcPPr-   & 1  & 2  & 0 &0      & 30 &60  & 1.00 & 5.5   \\
     SbcPPr+   & 1  & 2  & 0 &0      & 30 &60  & 1.00 & 44   \\
     SbcPol     & 1  & 2  & 90 &0      & 30 &60 & 1.00  & 11   \\
     SbcR      & 1  & 2  & 150 &0      & 150 &180  &0.60 & 50  \\
     \hline
     SbcPPn=0       & 1  & 0  & 0 &0      & 30 &60  & 1.00  & 11   \\
     SbcRn=0    & 1  & 0  & 150 &0      & 150 &180  & 0.60 & 50  \\
     \hline
     \multicolumn{9}{c}{G-G mergers} \\
     \hline
     G3PP        & 1  & 2  &-30 & 0     &30 & 60  &0.95  &13.6  \\
     G2PP        & 1  & 2  &-30 & 0     &30 & 60  &0.95  &3.8  \\
     G1PP        & 1  & 2  &-30 & 0     &30 & 60  &0.95  &3.0  \\
     G0PP        & 1  & 2  &-30 & 0     &30 & 60  &0.95  &2.2  \\ 
     \hline
     G3PPn=0        & 1  & 0  &-30 &0     &30 & 60  &0.95   &13.6  \\
     G2PPn=0        & 1  & 0  &-30 & 0     &30 & 60  &0.95  &3.8  \\
     G1PPn=0        & 1  & 0  &-30 & 0     &30 & 60  &0.95  &3.0  \\
     G0PPn=0        & 1  & 0  &-30 & 0     &30 & 60  &0.95  &2.2  \\ 
     \hline
   \end{tabular}
 \end{minipage}
\end{table} \label{simpartab}

Each of the galaxy merger simulations presented here are mergers of identical equal-mass galaxies;  mergers of
non-equal mass galaxies will be presented in a later paper.    The Sbc-Sbc merger simulation parameters are
selected to probe a range of different merger orientations and orbits (Table 2),   including parabolic orbits with 
roughly prograde-prograde (SbcPP), prograde-retrograde (SbcPR), retrograde-retrograde (SbcRR), and prograde-polar orientations (SbcPol), 
parabolic prograde-prograde mergers with very small/large
pericentric distances (SbcPPr-, SbcPPr+), and a highly radial orbit with prograde-retrograde orientation (SbcR).  
For most of the simulations, the galaxy orbits are initialized to be parabolic with the given pericentric distance; 
as the simulations progress, the galaxies lose angular momentum because of dynamical friction and eventually merge. 
The galaxies in the radial orbit simulation (SbcR) start out nearly at rest, so the pericentric distance and eccentricity
have little physical meaning in this case.
The G-G merger simulations were all run with the same orientation (prograde-prograde) 
and slightly sub-parabolic orbits (eccentricity $= 0.95$) to probe the effect of varying total mass (G3PP, G2PP, G1PP, G0PP; 
Table 2).  
Sub-parabolic orbits were chosen so that the lower mass mergers (G1, G0) would merge in less than 2 Gyr.  
This choice will affect the close pair time-scales, but is unlikely to affect the morphological disturbance 
time-scales which peak at the first pass and final merger (see \S 4.6 for discussion). 

The simulations only include feedback from supernovae.  Most of these Sbc simulations were run with the `stiff' $n=2$ supernova
feedback equation of state; the Sbc prograde-prograde merger and radial orbit merger were run with the isothermal $n=0$ supernova
feedback equation of state as well (SbcPPn=0, SbcRn=0).   To test the simulations for convergence, 
the Sbc prograde-prograde merger was also run with 
four and ten times as many particles as our typical simulations (SbcPPx4, SbcPPx10).   
All of the G-G simulations were run with both supernova feedback models. 
Although feedback from an active galactic nucleus (AGN) may be
important for the properties of the merger remnants,  such feedback is not expected to affect
the large-scale merger properties and morphology until after the coalescence of the nuclei.   Such feedback is driven by a rapidly accreting
AGN which may appear as an optically luminous quasar during the end stages of the merger. Incorporating a quasar into the
radiative transfer code poses a number of technical challenges and is beyond the scope of this paper, but will 
be included in a future version of SUNRISE.   As we discuss in
\S 5,  the exclusion of AGN feedback effects will not affect the morphological disturbance time-scales calculated here. 

\section{ANALYSIS}
We replicate the observations and measurements of real galaxy mergers as closely 
as possible.   Our simulations produce multi-wavelength images of galaxy mergers including the
effects of dust and star-formation.   Most current morphological measurements of the evolution 
of the galaxy merger rate are done in the rest-frame $B$ or SDSS $g$  
(e.g. Conselice et al. 2003, Lotz et al. 2008).  This is because       
high spatial resolution rest-frame $4000-5000$ \AA\ imaging can be done for local galaxies from the ground,  
for $z \sim 1$ galaxies with the {\it Hubble Space Telescope Advanced Camera for Surveys},  
and for $z \leq 3$ galaxies with the {\it HST Near-Infrared Camera} and {\it Wide Field Camera 3}.   
Therefore we focus on SDSS $g$ morphologies for purposes of this paper, as
these simulations can be used to calibrate the morphologies of galaxies currently observed at $0 < z < 3$. 
In the following section we describe how the simulated SDSS $g$ images are degraded and analysed to match real
galaxy morphology measurements. 

\subsection{Image degradation}
The SDSS $g$ images are produced by SUNRISE for each simulation for 11 isotropically positioned viewpoints as a function
of time from $\sim 0.5$ Gyr prior to the first pass to $\ge$ 1 Gyr after the final coalescence in $\sim 30-250$ 
Myr timesteps depending on the merger state.   The field of view of the output images ranges from 200 kpc 
during the initial stages and period of 
maximal separation to 100 kpc during the first pass, final merger and post-merger stages. 
The intrinsic resolution of the output SUNRISE $g$-band images is 333 pc.

The images output by SUNRISE have no background sky noise and no seeing effects, although they
do have particle noise and Monte Carlo Poisson noise. We degrade these images to simulate real data, 
but do not attempt to mimic a particular set of galaxy survey observations.  The measured morphology is dependent
on the observed spatial resolution and signal-to-noise to the extent that low spatial resolution ($>$ 500 pc per
resolution element) and low signal-to-noise (average $S/N$ per pixel $<$ 3-5) introduce biases in the morphology
values.   However, above these limits, measured morphologies are not dependent on spatial resolution or signal-to-noise
(LPM04).   
We re-bin the images to 105 pc per pixel and convolve the images with a Gaussian function with a FWHM = 400 pc.  
This was done to simulate the effect of seeing but maintain as high spatial resolution as possible.  The values where
chosen to match the SDSS with 1.5\arcsec\ seeing, 0.396\arcsec\ per pixel plate scale for a galaxy at a distance such
that 1.5\arcsec\ $\sim$ 400 pc.  We also add random Poisson noise background to simulate sky noise but scale this noise
to maintain a high $S/N$ for the galaxies ($>20$ per pixel within the Petrosian radius).  Our choice to simulate images
with spatial resolution $<$ 400 pc per resolution element and $S/N$ per pixel $>$ 20 means that our results here
can be generally compared to any rest-frame $\sim$ 4700 \AA\ morphological study with sufficient spatial resolution
and signal-to-noise.  This corresponds to galaxies with a distance modulus $<$ 35.0 observed from the ground with a seeing PSF
FWHM $\sim$ 0.8\arcsec  or galaxies at $z < 0.25$ observed with $HST$ $ACS$ and a PSF FWHM $\sim$ 0.14\arcsec .  In \S4.2 
we discuss how the results presented here apply to high-redshift galaxies observed with $HST$ at spatial resolutions $\sim$ 1 kpc,
where the morphological biases from spatial resolution can be important. 

\subsection{Morphology Measurements}
The degraded image for each snapshot and camera is treated as an independent observation with no prior information
except for the central position of the two galaxies, which is used to track the galaxies' identities.
Each image is run through an automated galaxy detection algorithm integrated into our IDL code.  This algorithm is 
similar to but simpler than the detection and de-blending algorithm of SExtractor (Bertins \& Arnouts 1996), 
which is optimized for large images with many objects but is ill-suited for images of one or two objects.  
The image is smoothed by a $\sim$ 4 kpc boxcar and initial segmentation maps of galaxies are determined based on a 
fixed surface-brightness threshold equal to 2$\sigma_{noise}$.  The number of objects detected in the 2$\sigma$ 
threshold map is compared to the number of objects detected in a 15$\sigma$ threshold map.   If more than 1 object 
larger than the smoothing length is detected in either map, a de-blending algorithm is applied.   The largest 
objects detected in the high threshold map are grown
using an image dilation algorithm (the IDL DILATE function) and a $5\times5$ pixel square shape operator, with a limiting
surface brightness set to the 2$\sigma$ threshold.   The de-blending algorithm adopted here results in similar segmentation maps
to those used in LPM04 for the sample of local mergers.

If the centres of the merging galaxies are less than 10 kpc apart, they are generally 
detected as  a single object. If 2 distinct galaxies are detected,  the detection segmentation maps are used to mask 
out the other galaxy while each galaxy's morphology is measured. The output segmentation maps are visually inspected.   
Occasionally, the detection algorithm will assign
a tidal dwarf galaxy as a second primary galaxy or fail to adequately mask out the secondary galaxy. In these
cases, the masking is done by hand and the morphology code is re-run.  For this paper, we ignore the 
properties of any tidal dwarfs produced in the merger and focus only on the merging galaxies and their remnants. 

The centres of each galaxy are estimated by minimizing the second-order moment of the pixels assigned to the
detection segmentation map, and ellipticities and position angles are determined using the IDL task FIT$\_$ELLIPSE
(Fanning 2002). The projected separation $R_{proj}$ is measured when two galaxies are detected. 
The initial guesses at the centre, ellipticity, and position angle are then used to calculate 
the Petrosian radii in circular and elliptical apertures, concentration, asymmetry, clumpiness, 
the Gini coefficient, and the second-order moment of the brightest 20\% of the light 
(see Lotz et al. 2004 and Conselice 2003 for further details).   

The Petrosian radius is defined as the radius $r_p$ at which 
the ratio of the surface brightness at $r_p$ to the mean surface brightness within $r_p$ is equal
to a fixed value, i.e.
\begin{equation}
\eta = \frac{\mu(r_p)}{\bar{\mu}(r<r_p)}
\end{equation}
where $\eta$ is typically set to 0.2 (Petrosian 1976).  
Because the Petrosian radius is based on a curve of growth, it is largely insensitive 
to variations in the limiting surface brightness and  $S/N$ of the observations.

Concentration is defined in slightly different ways by different authors, but the basic
function measures the ratio of light within a circular inner aperture 
to the light within an outer aperture.  We adopt the Bershady et al. (2000) definition as the ratio of the 
circular radii containing 20\% and 80\% of the ``total flux'' :
\begin{equation}
C = 5\ \rm{ log10}\left(\frac{r_{80}}{r_{20}}\right)
\end{equation}
where $r_{80}$ and $r_{20}$ are the circular apertures containing 80\% and 20\%
of the total flux, respectively.   For comparison to the most recent
studies of galaxy concentration, we use Conselice's (2003) definition of the total flux as
the flux contained within 1.5 $r_p$ of the galaxy's centre (as opposed to 
Bershady's definition as the flux contained within 2 $r_p$).  For 
the concentration measurement, the galaxy's centre is that determined by
the asymmetry minimization (see below).  Bulge-dominated early-type galaxies
generally have high concentrations (C $\sim$ 4-6), while late-type discs have low concentrations (C $\sim$ 2-3).
On-going mergers may show very low concentrations or very high concentrations depending
on the merger stage and the brightness of the central starburst.

The asymmetry parameter $A$ quantifies the degree to which the light of a galaxy is 
rotationally symmetric.  
$A$ is measured by subtracting the galaxy image rotated by 180 degrees from the
original image (Abraham et al. 1994, Conselice et al. 2000). 
\begin{equation}
A = \sum_{i,j} \frac{ | I(i,j) - I_{180}(i,j)|}{|I(i,j)|} - B_{180}
\end{equation}
where $I$ is the galaxy's image and $I_{180}$ is the image rotated by 180 about
the galaxy's central pixel, and $B_{180}$ is the average asymmetry of the background. 
$A$ is summed over all pixels within 1.5 $r_{p}$ of the galaxy's centre.  The
central pixel is determined by minimizing $A$.  The asymmetry due to the noise must be 
corrected for, and it is impossible to reliably 
measure the asymmetry for very low signal-to-noise ratio images (LPM04).
Objects with very smooth elliptical light profiles have low asymmetries (A $< 0.05$). 
Galaxies with spiral arms are more asymmetric (A $\sim$ 0.1-0.2), while extremely irregular and 
merging galaxies are often highly asymmetric (A $>$ 0.35).     

The smoothness parameter $S$ has been developed by Conselice (2003), inspired by
the work of Takamiya (1999), in order to quantify the degree of small-scale structure.  
The galaxy image is smoothed by a boxcar of given width and then
subtracted from the original image.   The residual is a measure of the clumpiness due to
features such as compact star clusters.  
In practice, the smoothing scalelength is chosen to be a fraction of the Petrosian radius.
\begin{equation}
S =  \sum_{i,j}\frac{ | I(i,j) - I_S(i,j)| } {|I(i,j)|} - B_S 
\end{equation}
where $I_S$ is the galaxy's image smoothed by a boxcar of width 0.25 $r_p$, and
$B_S$ is the average smoothness of the background.  Like $A$, $S$ is summed over the
pixels within 1.5 $r_p$ of the galaxy's centre.  However, because the central regions
of most galaxies are highly concentrated, the pixels within a circular aperture
equal to the smoothing length 0.25 $r_p$ are excluded from the sum.
$S$ is correlated with recent star-formation (Conselice 2003).

The Gini coefficient $G$ is a statistic based on the Lorenz curve, the 
rank-ordered cumulative distribution function of a population's wealth or, in this
case, a galaxy's pixel values (Abraham et al. 2003).  
The Lorenz curve is defined as 
\begin{equation}
L(p) = \frac{1}{\bar{X}} \int_0^p{F^{-1}(u) du}
\end{equation}
where $p$ is the percentage of the poorest citizens or faintest pixels, 
F(x) is the cumulative distribution function, and $\bar{X}$ is the
mean over all (pixel flux) values $X_i$ (Lorenz 1905).   
The Gini coefficient is the ratio of the area between 
the Lorenz curve and the curve of ``uniform equality'' (where $L(p) = p$) 
to the area under the curve of uniform equality ($= 1/2$). 
For a discrete population, the Gini coefficient 
is defined as the mean of the absolute difference between all $X_i$:
\begin{equation}
G = \frac{1}{2 \bar{X} n (n-1)} \sum^n_{i=1} \sum^n_{j=1} | X_i - X_j |
\end{equation}
where $n$ is the number of people in a population or pixels in a galaxy. 
In a completely egalitarian society, $G$ is zero, and if one individual has all the wealth, $G$ is unity.  
A more efficient way to compute $G$ is to first sort $X_i$ into increasing order
and calculate
\begin{equation}
G = \frac{1}{\bar{|X|} n (n-1)} \sum^n_i (2i - n -1) |X_i|
\end{equation}
(Glasser 1962).
$G$ is high for objects with very bright nuclei (G $\sim$ 0.6), whether those galaxies are highly concentrated
ellipticals or mergers with multiple bright nuclei.  It is low for objects with more uniform
surface brightnesses, such as late-type discs (G $\sim$ 0.4). 

Because $G$ is very sensitive to the ratio of low surface brightness to high surface
brightness pixels, $G$ must be measured within a well-defined segmentation map.  
For the purposes of measuring $G$ and $M_{20}$, we have chosen to assign pixels to the galaxy 
based on the surface-brightness at the Petrosian radius as measured in elliptical apertures
(LPM04; see Abraham et al. 2007 for a similar approach).  Note that the resulting segmentation map is
not elliptical, but rather traces the isophote that matches the mean surface-brightness at the
elliptical Petrosian radius.  The Petrosian radius is a reproducible quantity that is relatively independent of 
signal-to-noise and surface-brightness dimming effects (LPM04).   Designating galaxy pixels based on
signal-to-noise cuts (e.g. Law et al. 2007) will result in unreliable $G$ values. 
This is because galaxies with the same `shape' or `morphology' but differing luminosities will have different
measured $G$ values when signal-to-noise is used to define the galaxy pixels.  For example, a low
surface-brightness exponential disc will have fewer low surface-brightness pixels assigned
to its segmentation map than a morphologically identical high 
surface-brightness exponential disc, resulting in a lower $G$ for the low surface-brightness disc. 
Moreover, $G$ values measured within segmentation maps based on signal-to-noise cuts are 
not repeatable because the measured $G$ will depend as much on the noise properties of the observations 
as the intrinsic galaxy properties.  The results presented in this work are robust to these effects because  
the pixel maps used to measure $G$ are based on the surface brightness at the Petrosian radius. 

The total second-order moment $M_{tot}$ 
is the flux in each pixel, $f_i$, multiplied by the squared distance to the centre of the galaxy, 
summed over all the galaxy pixels assigned by the segmentation map:
\begin{equation} 
M_{tot} = \sum_i^n M_i = \sum_i^n f_i \cdot ((x_i - x_c)^2 + (y_i - y_c)^2)
\end{equation}
where $x_c, y_c$ is the galaxy's centre. The centre is computed by finding $x_c, y_c$ 
such that $M_{tot}$ is minimized.
The second-order moment of the brightest regions of the galaxy  
traces the spatial distribution of any bright nuclei, bars, 
spiral arms, and off-centre star-clusters. 
$M_{20}$ is defined as the normalized second order moment of the brightest 20\% of
the galaxy's flux.  To compute $M_{20}$, we rank-order the galaxy pixels by flux, 
sum $M_i$ over the brightest pixels until the sum of the brightest pixels equals 20\% of the 
total galaxy flux, and then normalize by $M_{tot}$:
\begin{eqnarray}
M_{20} \equiv {\rm log_{10}}\left(\frac{\sum_i M_i}{M_{tot}}\right) & {\rm while } & \sum_i f_i <  0.2 f_{tot}
\end{eqnarray}
Here $f_{tot}$ is the total flux of the galaxy pixels identified by the
segmentation map and $f_i$ are the fluxes for each pixel $i$, ordered such that $f_1$ is the brightest pixel, 
$f_2$ is the second brightest pixels, and so on. The normalization by $M_{tot}$ 
removes the dependence on total galaxy flux or size. $M_{20}$ always has
a value $<$ 0.  $M_{20}$ is anti-correlated with $C$ for
normal galaxies, with low values for early-type galaxies ($M_{20} \le -2$) and intermediate values
for late-type galaxies ($M_{20} \sim -1.5$).   Mergers with multiple nuclei have high $M_{20}$ values
( $\ge -1$). 

\subsection{Definition of Merger Stages}
From the true three-dimensional separations of the galaxy nuclei, 
we determine the timestep of the closest approach during the first pass ($t_{fp}$), 
maximal separation after the first pass ($t_{max}$), and final merger of the nuclei where $\delta$r $<$ 1 kpc ($t_{merg}$).  
These times are given in Table \ref{stagetab} for each simulation.  
Based on these events, we define 6 merger stages: pre-merger,  first pass,  
maximal separation, final merger, post-merger, and merger remnant. 
The `pre-merger' stage is from t=0 to 0.5 $t_{fp}$. 
The `first pass' stage encompasses the first pass and starts at 0.5 $t_{fp}$ and ends at 0.5 ($t_{fp} + t_{max}$). 
The `maximal separation' stage starts at  0.5 ($t_{fp} + t_{max}$) and ends at 0.5 ($t_{max} + t_{merg}$). 
The `merger' stage starts at  0.5 ($t_{max} + t_{merg}$) and ends at $t_{merg} +$ 0.5 Gyr.  
The `post-merger' stage is defined as between $t_{merg} +$ 0.5 Gyr and $t_{merg} +$ 1.0 Gyr,  
while the `remnant' stage is at times more than 1 Gyr after the merger event ($t_{merg} +$ 1.0 Gyr).
We show composite SDSS $u-r-z$ images for each of these stages
for a prograde-prograde Sbc merger simulation as viewed face-on (Figure 1) and edge-on (Figure 2). 


\begin{table}
  \begin{minipage}{84mm}
  \caption{Merger Stages}
  \begin{tabular}{@{}lccccc@{}}
    \hline
    Simulation & First Pass & Max. Sep.  & Merger & Post-Merger  &Remnant \\
    & (Gyr)  &(Gyr) & (Gyr) & (Gyr)  & (Gyr) \\
    \hline
    \multicolumn{6}{c}{Sbc-Sbc mergers} \\
    \hline
    SbcPPx10  & 0.59  & 1.03  & 1.66  & 2.16 & 2.66 \\
    SbcPPx4   & 0.59  & 1.03  & 1.71  & 2.21 & 2.71 \\
    \hline
    SbcPP     & 0.59  & 1.03  & 1.71  & 2.21 & 2.71 \\
    SbcPR     & 0.59  & 1.08  & 1.71  & 2.21 & 2.71 \\     
    SbcRR     & 0.59  & 1.12  & 1.71  & 2.21 & 2.71 \\
    SbcPPr-   & 0.59  & 0.88  & 1.37  & 1.87 & 2.37 \\
    SbcPPr+   & 0.68  & 1.91  & 3.76  & 4.26 & 4.76 \\
    SbcPol     & 0.59  & 1.17  & 2.00  & 2.50 & 3.00 \\
    SbcR      & 1.28  & 1.47  & 1.70  & 2.20 & 2.70 \\
    \hline
    SbcPPn=0    & 0.59  & 1.03  & 1.71  & 2.21 & 2.71 \\
    SbcRn=0    & 1.28  & 1.42  & 1.66  & 2.16 & 2.66 \\
    \hline
    \multicolumn{6}{c}{G-G mergers} \\
    \hline
    G3PP       & 0.85  & 1.47  & 2.44  & 2.93 & 3.43 \\     
    G2PP        & 0.40  & 0.70  & 1.24  & 1.74 & 2.24 \\ 
    G1PP        & 0.45  & 0.68  & 1.24  & 1.74 & 2.24 \\
    G0PP        & 0.55  & 0.88  & 1.42  & 1.92 & 2.41 \\  
    \hline
    G3PPn=0     & 0.83  & 1.37  & 2.40  & 2.90 & 3.40 \\    
    G2PPn=0     & 0.39  & 0.68  & 1.17  & 1.67 & 2.17 \\
    G1PPn=0     & 0.45  & 0.68  & 1.24  & 1.74 & 2.24 \\
    G0PPn=0     & 0.54  & 0.83  & 1.32  & 1.82 & 2.32 \\
    \hline
  \end{tabular}
  \medskip \\
  The time since the start of the simulation is given for 
  each event that defines a particular merger stage (see \S 3.3 for definitions). 
  \end{minipage}
\end{table} \label{stagetab}

\subsection{Merger Classification and Time-Scales}

\begin{figure*}
\includegraphics[width=160mm]{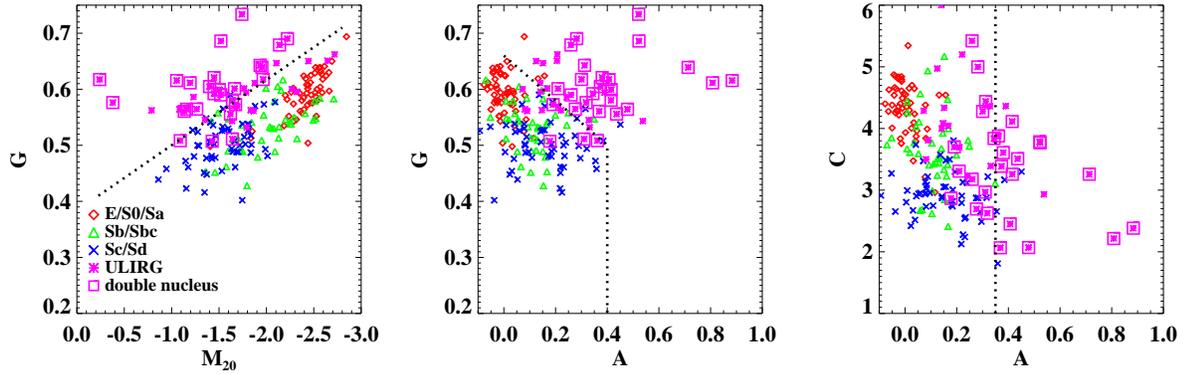}
\caption{$G-M_{20}$, $G-A$, and $C-A$ relations for local galaxies measured by
LPM04.   Empirically, normal galaxies lie below the dashed lines in the $G-M_{20}$ and $G-A$ plots and to
the left of the dashed line in the $C-A$ plot. Mergers are identified as galaxies which lie above and/or to the right
of these divisions.  While 90\% of local ULIRGs show visual signs of merger activity,  ULIRGs with double nuclei show the strongest
signatures in quantitative morphology.} \label{gm20_local}
\end{figure*}

\begin{figure*}
\includegraphics[width=160mm]{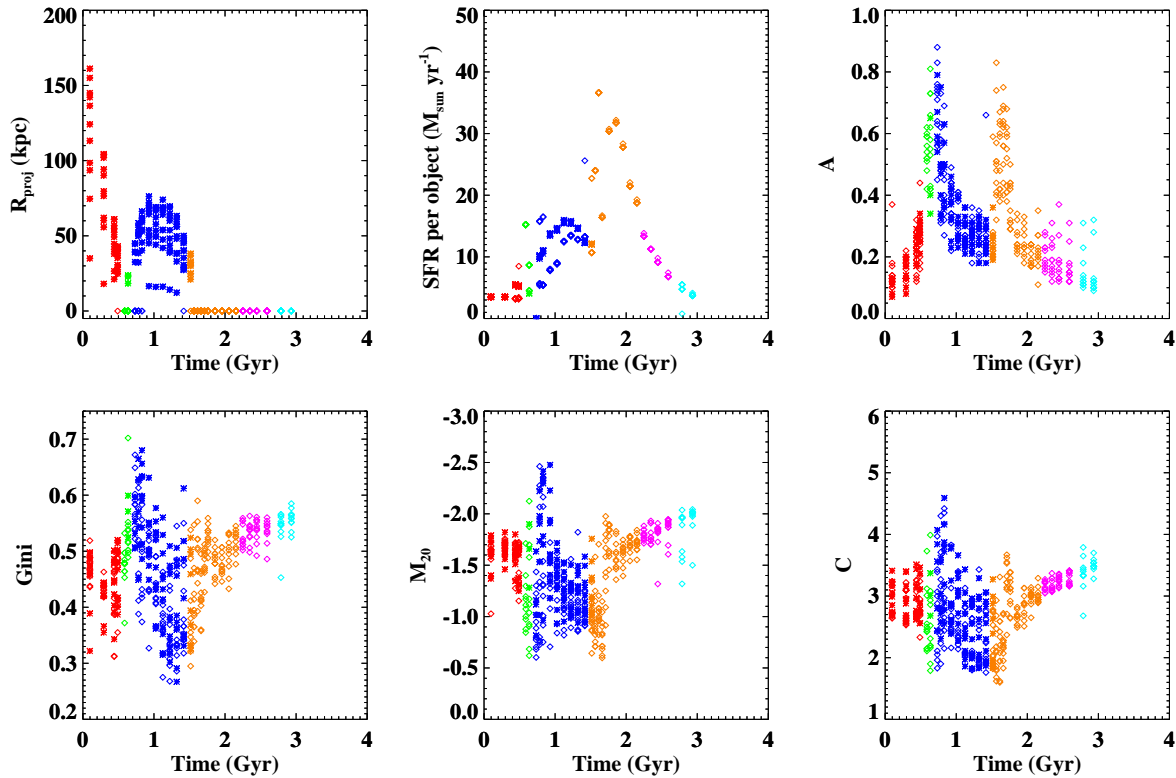}
\caption{Time v. R$_{proj}$, star-formation rate per object, $A$, $G$, $M_{20}$, and $C$ for the
high-resolution prograde-prograde Sbc merger with stiff supernova feedback, including the effects of dust 
(SbcPPx10). Each merger stage is marked with a different colour
(pre-merger: red,  first pass: green, maximal separation: blue, final merger: orange, post-merger: magenta,
remnant: cyan).  Open diamonds are for the one merging galaxy and the merger remnant; crosses are for the other merging galaxy. The mergers show strong morphological disturbances and peaks in the star-formation rate
 at the first pass (green points) and final merger (orange points).  }\label{timeSbc201a10x}
\end{figure*}

In Figure \ref{gm20_local}, we show the empirical criteria for merger classification via $G-M_{20}$, $G-A$ 
and $C-A$ morphologies for local samples of visually classified galaxies (LPM04). All three 
morphological merger criteria are based on the Borne et al. (2000) observations of local ultra-luminous
infrared galaxies (ULIRGs), a subset of which have been shown to be gas-rich mergers with mass ratios between 1:3 and 1:1
(Dasyra et al. 2006).   LPM04 found that ULIRGs visually classified as mergers could be distinguished from
the sequence of normal Hubble type galaxies with  
\begin{equation}
G > -0.115\ M_{20} + 0.384
\end{equation}
or
\begin{equation}
G  > -0.4\ A + 0.66\ {\rm or}\  A \ge 0.4
\end{equation}
Asymmetry alone is also often used to classify merger candidates.   The calibration of local mergers by
Conselice (2003) finds the following merger criterion:
\begin{equation}
A \ge 0.35
\end{equation} 
These are the merger criteria used to calculate the morphological observability time-scales throughout most of this paper. 

Galaxies at higher redshift cannot be imaged at as high spatial resolution as local galaxies even
when observed with $HST$.  The measured morphologies of galaxies at $z > 0.25$ imaged with $HST$ will have
non-negligible biases as a result of this lower spatial resolution (LPM04). Therefore the merger criteria have been adjusted
to account for these biases in $HST$ data by Conselice et al. (2003), Conselice, Blackburne, \& Papovich (2005), and Lotz et al. (2008).  
For $z < 1.2$ galaxies observed with $HST$ $ACS$ Wide-Field Camera at rest-frame $\sim$ 4000\AA, the
revised $G-M_{20}$ merger classification is:
\begin{equation}
G > -0.14\ M_{20} + 0.33
\end{equation}
(Lotz et al. 2008). Conselice et al. (2005) find that decreased spatial resolution and surface brightness dimming
at $z>0.5$ can lower the measured $A$ in irregular galaxies by 0.05-0.15 (also Shi 2008).  In \S4.2, we find a median offset of
$-0.05$ for $A$ when our simulations are convolved to match the spatial resolution of $HST$ $ACS$ $WFC$ $F814W$ observations
at $z \sim 1$.   Therefore we suggest a revised merger criterion for $G-A$ and $A$ for $HST$ observations of high-redshift galaxies 
as follows:
\begin{equation}
G > -0.4\ A + 0.68\ {\rm or}\  A \ge 0.35
\end{equation}
and
\begin{equation}
A \ge 0.30
\end{equation}
We will use these merger criteria in \S4.2 for the simulations convolved to match $HST$ $ACS$ observations of galaxies at $z \sim 1$.

Close kinematic pairs are also probable merging systems.  Recent studies of local kinematic 
pairs have selected objects with $5 < R_{proj} < 20 h^{-1}$ kpc (Patton et al. 2002, de Propris et al. 2005)
and relative velocities $\leq 500$ km s$^{-1}$,  
while studies of pairs out to $z \sim 1.4$ has chosen objects with $10 < R_{proj} <$ 30, 50, 
and 100 $h^{-1}$ kpc and relative velocities $\leq$ 500 km s$^{-1}$ (Lin et al. 2004).   
We assume $h=0.7$ and we estimate the time-scales during which merging galaxies can be found as separate objects
within $5  < R_{proj} < $ 20, 10 $ < R_{proj} <$ 30, 10 $ < R_{proj} < $ 50, 
and 10 $< R_{proj} < $ 100 $h^{-1}$ kpc.    The simulated merging galaxies always have relative velocities $<$ 500  
km s$^{-1}$. 

The galaxy merger rate $\Gamma$ is defined as the number of on-going merger events per unit volume
$\phi_{merg}$ divided by the time $T_{merg}$ for the merger to occur from the initial encounter to the final coalescence:
\begin{equation}
\Gamma = \frac{\phi_{merg}}{T_{merg}}
\end{equation}
However, the number density of galaxies identified morphologically as galaxy mergers $\phi^{'}_{merg}$ will depend on the 
time-scale  $T^{'}_{merg}$ during which the merger can be observed given the method used to identify it such
that
\begin{equation}
\phi^{'}_{merg} = \phi_{merg} \frac{T^{'}_{merg}}{T_{merg}}
\end{equation}
Therefore the galaxy merger rate $\Gamma$ can be calculated from the observed number density of galaxy merger candidates
$\phi^{'}_{merg}$ as follows:
\begin{equation}\label{merg_eqn}
\Gamma = \frac{\phi^{'}_{merg}}{T_{merg}} \frac{T_{merg}}{T^{'}_{merg}} =  \frac{\phi^{'}_{merg}}{T^{'}_{merg}}
\end{equation}

The effective observability time-scale $T^{'}$ given in Equation \ref{merg_eqn} is a weighted average of the time-scales over all likely
merger mass ratios, gas fractions, and orbital parameters.  We do not calculate this global observability time-scale
here because this may require a cosmological model for the distribution of galaxy merger properties if the observability
time-scale for a given method varies strongly.  
Instead, we present the first steps towards calculating the global observability time-scales by exploring the 
the dependence of the time-scales on a wide range of galaxy merger properties. 

We calculate each simulation's average observability time-scale for the $G-M_{20}$, $G-A$, and $A$ criteria given
above by averaging the results of the 11 isotropic viewpoints.  Hereafter we drop the prime notation and refer to the
observability time-scale for each simulation as $T$. 
Because we wish to determine the number density of merger events rather than the number of galaxies
undergoing a merger, galaxies that have not yet merged but identified morphologically as merger candidates are 
weighted accordingly.  The time that each pre-merger galaxy is morphologically disturbed is summed (not averaged)
to the time that the post-merger system appears disturbed. 
No such weighting is done for the close pair time-scales as this factor is generally
included in the merger rate calculation (e.g. Patton et al. 2000):
\begin{equation}
\Gamma = 0.5\ \phi\ N_c\ p(merg)\ T_{pair}^{-1}
\end{equation}
where $\phi$ is the number density of galaxies within the magnitude range of the observed pairs, $N_c$ is the
average number of companions within the observed magnitude range per galaxy, $p(merg)$ is the probability that a galaxy pair
will merge, $T_{pair}$ is the time-scale for which merging galaxies will meet the close pair criteria, and 0.5 is the weighting
factor that accounts for the double counting of pairs.

\begin{figure*}
\includegraphics[width=168mm]{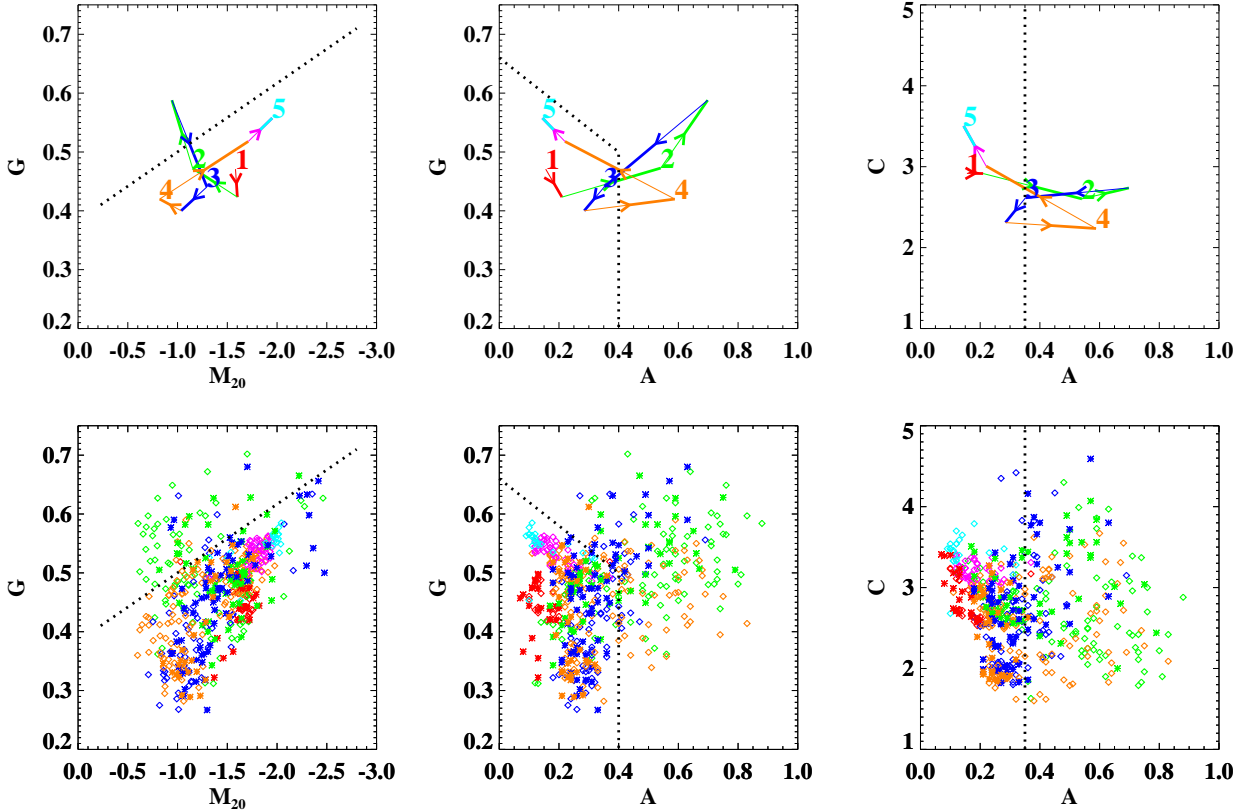}
\caption{Top: Morphological evolutionary tracks in $G-M_{20}$, $G-A$, and $C-A$ space 
for the high resolution prograde-prograde Sbc merger (SbcPPx10) averaged over all viewing angles.  
The initial galaxy (red), first pass (green), maximal separation (blue), final merger (orange), 
and last computed merger remnant (cyan) morphologies are labeled as 1-5 respectively.
The black dashed lines show the empirical merger classification criteria from Figure 3. 
Bottom: The morphologies for all the timesteps
and cameras for the SbcPPx10 simulation.  The merger stages are indicted by the different colours, as
in Figure \ref{timeSbc201a10x}. For this simulation, $G-M_{20}$ identifies SbcPPx10 mergers primarily at the first pass,
while $G-A$ and $A$ find mergers at both the first pass and final merger.}\label{mergertracks}
\end{figure*}

\section{Results}
The equal-mass galaxy merger simulations span a wide range of physical properties.  These include the
relative orientations and orbital parameters of the merging galaxies, the gas fraction and scalelength,
the assumed supernova feedback prescription, and masses.   In this section, we explore how important 
these physical parameters, the inclusion of dust, and simulation resolution are to the predicted morphologies.  
We present the viewing-angle averaged observability time-scales for $G-M_{20}$, $G-A$, and $A$ morphologies (Table \ref{timetab}) and
close pair projected separations $R_{proj} < 20, 30, 50, $ and 100 $h^{-1}$ kpc for each simulation (Table \ref{timetab2}).   
Finally,  the morphologies of the remnants observed 1 Gyr after the merger 
are calculated (Table \ref{remtab}). 

For each simulation, we examine the projected separations $R_{proj}$, measured morphologies ($G$, $M_{20}$, $C$, $A$, and $S$), 
and star-formation rate per object as a function of time and merger stage for each simulation for all 11 viewpoints.
The initial segmentation maps computed to identify each galaxy are used to compute the total star-formation rate for each galaxy
at each timestep and camera. 
In general we only examine the morphologies of the output images that include the effects of dust absorption and scattering.
In Figure \ref{timeSbc201a10x}, we show the evolution with time for the highest resolution prograde-prograde
Sbc merger simulation, SbcPPx10. The
merger stages are colour-coded, with red for pre-merger, green for first pass, blue for maximal separation, 
orange for merger, magenta for post-merger, and cyan for the remnant.  The scatter at each timestep reflects the
scatter in morphology with viewing angle. Before the final merger, the morphologies are measured separately for each galaxy (open
diamonds and crosses).  After the merger, the system is treated as a single object (open diamonds). 

Most of the parabolic-orbit mergers show trends of morphology and star-formation rate 
with merger stage similar to the SbcPPx10 merger (Figures \ref{timeSbc201a10x} 
and \ref{mergertracks}). Leading up to and including the first pass
(green points), the morphologies become increasingly disturbed as tidal tails are formed and the galaxies overlap in projection. 
The star-formation rate per object peaks at the first pass because the system appears as one object, but remains enhanced
above the initial star-formation rates as the galaxies approach their maximal separations (blue points).  Dust starts to obscure
star-formation in the nuclei, lowering $G$ and $C$ and increasing $M_{20}$.  Strong morphological disturbances are observed again
at the final merger (orange points).  The star-formation rate per object reaches its peak at or just after the final merger 
and generally continues at high levels until $\sim$ 0.5-1 Gyr after the coalescence of the nuclei. 
In Figure \ref{mergertracks}, we show the progression of the SbcPPx10 merger in $G-M_{20}$, $G-A$, and $C-A$ averaged over all viewing angles
(top) and for all 11 cameras (bottom).   The initial galaxies (red points) start with late-type disc morphologies in all three plots. 
The system become disturbed in $G-M_{20}$ space during the first pass (green points) and disturbed in $G-A$ and $C-A$ during the first pass
and final merger (orange points).   The post-mergers and remnants gradually end up with early-type disc morphologies in all three plots (cyan
points). 

\subsection{Numerical Resolution}


\begin{table}
    \caption{Morphological Timescales v. Resolution}
    \begin{tabular}{lcccc}
      \hline
      Simulation & T($G-M_{20}$)  & T($G-A$) & T($A$) \\
                 & (Gyr)          & (Gyr)  & (Gyr) \\
      \hline
      \multicolumn{4}{c}{No corrections} \\
      \hline
      SbcPPx10  &  0.27$\pm$ 0.13 &  0.95$\pm$ 0.16 &  1.01$\pm$ 0.14   \\
      SbcPPx4   &  0.30$\pm$ 0.12 &  0.79$\pm$ 0.17 &  0.82$\pm$ 0.17   \\
      SbcPP     &  0.77$\pm$ 0.15 &  2.12$\pm$ 0.51 &  2.74$\pm$ 0.30   \\
      \hline
      \multicolumn{4}{c}{t $\ge$ 0.6 Gyr} \\
      \hline
      SbcPPx10 &  0.26$\pm$ 0.10 &  0.90$\pm$ 0.14 &  0.94$\pm$ 0.13   \\
      SbcPPx4  &  0.30$\pm$ 0.11 &  0.78$\pm$ 0.17 &  0.80$\pm$ 0.15   \\
      SbcPP    &  0.56$\pm$ 0.19 &  1.45$\pm$ 0.31 &  1.94$\pm$ 0.28   \\
      \hline
      \multicolumn{4}{c}{t $\ge$ 0.6 Gyr; $\delta M_{20} = -0.157$;  $\delta A = -0.115$}\\
      \hline
      SbcPP &  0.39$\pm$ 0.16 &  0.78$\pm$ 0.21 &  0.74$\pm$ 0.17   \\
      \hline
      \end{tabular} \label{reslntab}
      \medskip \\
      PP=prograde-prograde;  x10 = 10 N$_{part}$;  x4 = 4 N$_{part}$
\end{table}

\begin{figure*}
\includegraphics[width=168mm]{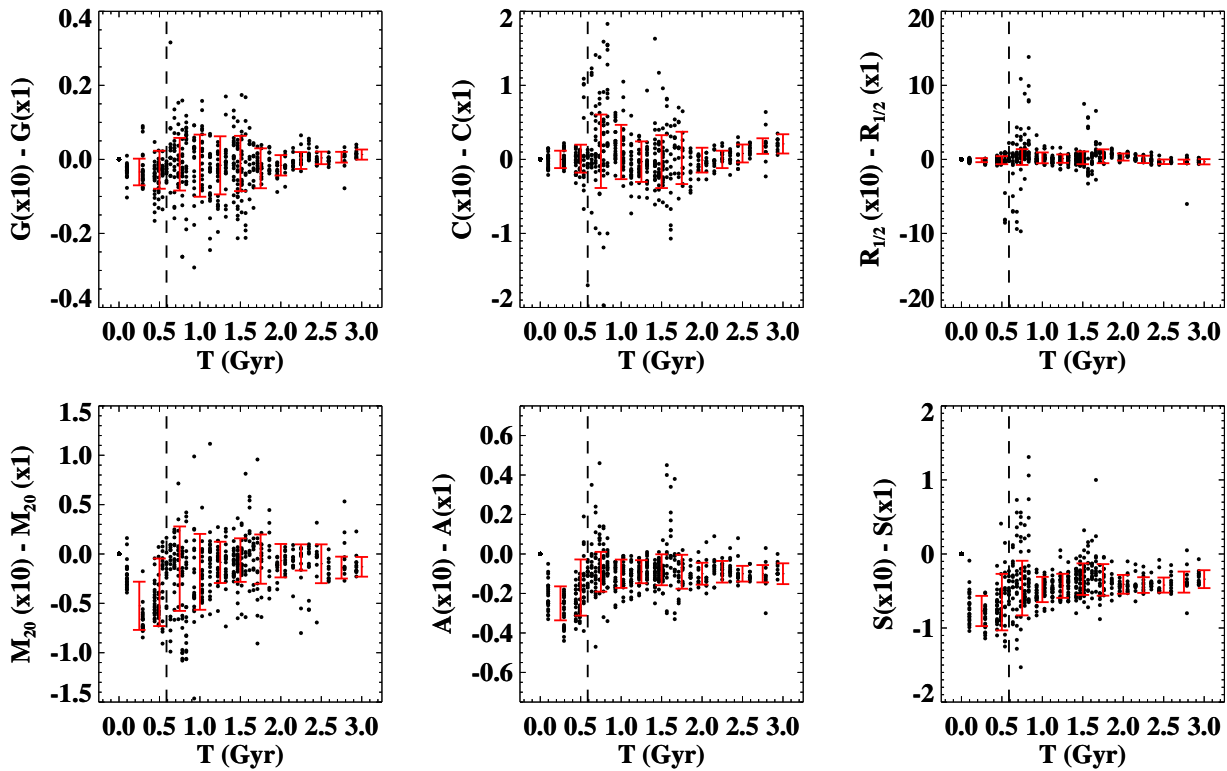}
\caption{$\Delta$ Morphology v. Time 
for the prograde-prograde Sbc simulation run with 10$\times$ (SbcPPx10) and 1$\times$ (SbcPP) the
standard number of particles. The red error-bars show the standard deviation of the morphology differences 
within 0.25 Gyr bins. Prior to 0.6 Gyr (dashed line), the low-resolution simulations has higher $M_{20}$ and $A$
values than the high-resolution simulation.}\label{dif10x} 
\end{figure*}

\begin{figure} 
\includegraphics[width=84mm]{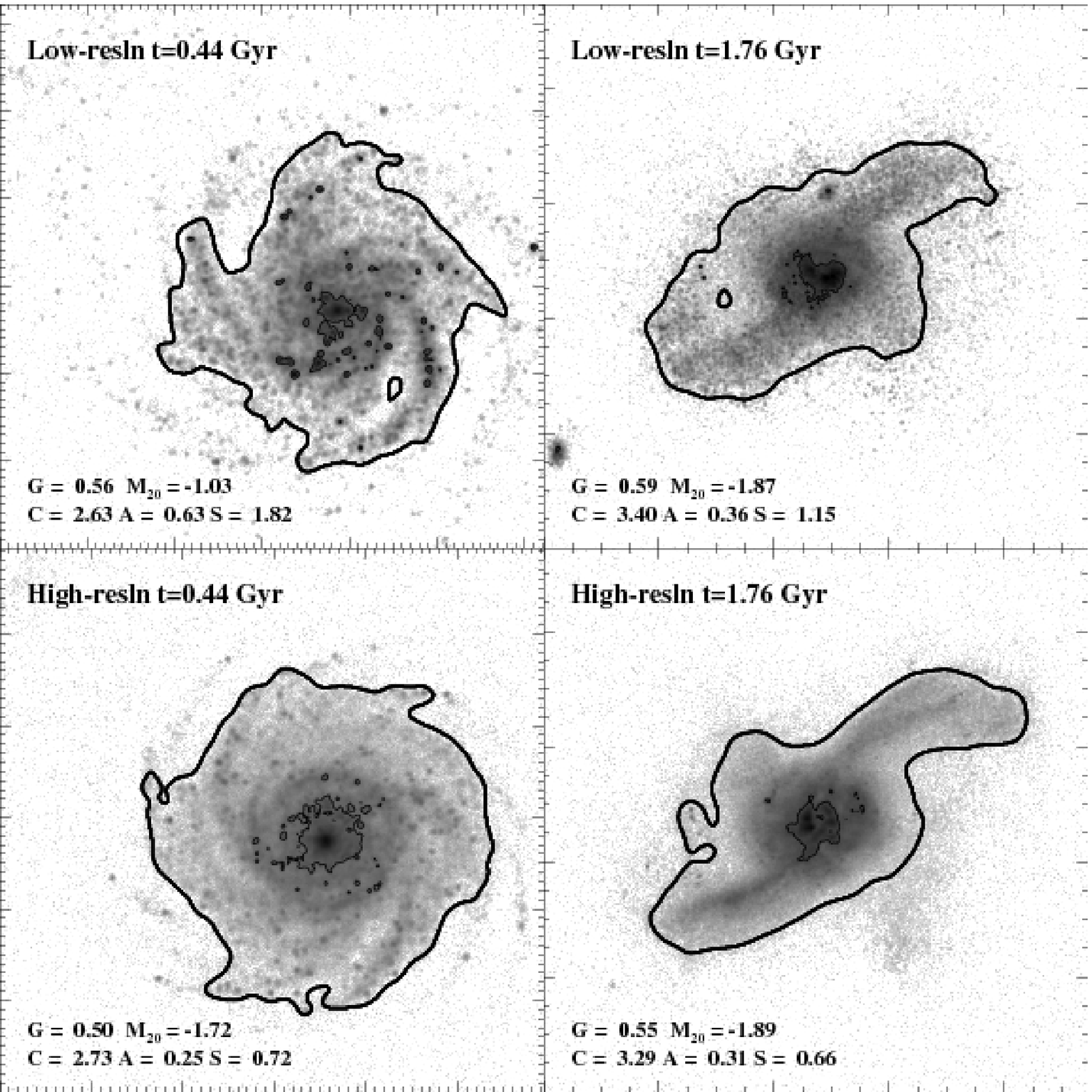}
\caption{Top: SDSS g-band images for standard resolution simulation SbcPP at time = 0.44 Gyr and
1.76 Gyr.   Bottom: Same for high numerical resolution SbcPPx10 simulation.   Thick contours show the
segmentation maps used to compute $G$ and $M_{20}$.  The thin contours show the pixels containing
brightest 20\% flux.   Prior to the merger, the relatively low star-formation surface densities result in new stars that
are concentrated in a single particle for a given star-forming region in the standard
resolution simulations, resulting in artificially high $M_{20}$, $A$, and $S$ values.  During and after the merger, 
high star-formation surface densities produce multiple new star particles per star-forming region, resulting in
more consistent morphologies between the standard and high numerical resolution simulations.}\label{reslnimg}
\end{figure}

Most of our simulations were run with $\sim 10^5$ particles per galaxy, with gravitational softening lengths of 400 pc 
and 100 pc for the dark matter and baryonic particles respectively (Table 1).  The number of simulation particles
affects both the spatial resolution of the simulation and the noise in the fluctuations of the gravitational
potential.  This number of particles was found to adequately recover the system-averaged
star-formation histories and remnant properties when compared to simulations with larger numbers of particles (Cox et al. 2006). 
Here we compare the time-dependent morphologies of the stiff supernova feedback prograde-prograde Sbc merger (SbcPP)
to simulations run with 4 and 10 times as many particles (SbcPPx4 and SbcPPx10) to determine if the standard numerical
resolution of the simulations is adequate also for analysing morphology. 
These higher numerical resolution simulations were processed by SUNRISE and the morphologies
of the output SDSS $g$ images were compared to the fiducial simulation at each timestep.   
The time evolution is slightly different in distinct simulation runs, so when the morphologies are rapidly changing, 
the morphologies, half-light radii ($R_{1/2}$), and 
galaxy separations may be significantly different for the different runs at a given timestep.
However, for the majority of the simulation timesteps, differences
in the morphologies will reflect the differing spatial resolution and the noise in the gravitational potentials of the simulations.
We find no resolution dependence for the time-scales of close pair projected separations $R_{proj}$, and so
we focus only on the morphologies in this section. 

In Figure~\ref{dif10x}, we plot the difference between the standard resolution simulation SbcPP
and the highest resolution simulation SbcPPx10 as a function of time for all 11 viewing angles 
including the effects of dust.   We find strong offsets in the half-light radii ($R_{1/2}$) and 
morphologies for a few timesteps immediately after the first pass at 0.6 Gyr and immediately before 
the final merger at 1.7 Gyr,  as expected from small timing differences between the different simulations.  
For the majority of timesteps, the mean differences between $G$, $C$, and effective radii for the standard and high 
resolution simulations are negligible but show significant scatter with viewing angle after the first pass.  
$M_{20}$, $A$ and $S$ do show systematic offsets  particularly for the initial undisturbed galaxies (t $<$ 0.6 Gyr;
dashed vertical line in Figure~\ref{dif10x}).  

We compare the location of the brightest 20\% of the 
pixels for the standard resolution and high resolution images of the initial discs in Figure~\ref{reslnimg}. 
Young star particles in the spiral arms of the initial galaxies dominate the morphologies
because they are not adequately sampled in the fiducial simulations. Because star formation is implemented by creating star
particles whose mass depends on the resolution of the simulation,  the fiducial simulations have fewer and brighter star clusters. 
For isolated and pre-merger galaxies,  most new star particle formation happens at star-formation surface densities 
close to resolution of the output images ($\Sigma_{SFR} \sim 0.03 \Msun$ yr$^{-1}$ kpc$^{-2}$, for $10^5$ $\Msun$ new star particles, 
images with a spatial resolution $\sim$ 400 pc, and O/B star lifetimes $\sim$ 20 Myr; see also Figure 3 in Cox et al. 2006).   
Therefore the new stars are concentrated into
a single particle within a single spatial resolution element, and the stochastic formation of individual star particles has a strong influence on the
morphologies. As the numerical resolution is increased and the mass of a new star particle decreases, 
new stars are distributed over several particles.  This decreases the typical surface brightness of the star-forming regions and the morphologies are less
dominated by stochastic star particle formation.  During the merger and remnant stages,  most of the star-formation happens at gas densities
well above the numerical limit of the fiducial simulations (see Figure 8 in Cox et al. 2006). 
Therefore the new stars are already distributed over multiple particles for the fiducial
simulations, and the morphologies are not dominated by stochastic new star particle formation (Figure~\ref{reslnimg}).

After the first pass, the standard resolution simulation continues to show small but significant
offsets to higher $M_{20}$, $A$, and $S$ relative to the high resolution simulation (Figure~\ref{dif10x}). 
Even when we ignore the $t < 0.6$ Gyr timesteps, we still find morphological disturbance time-scales twice as long 
for the standard resolution simulations (Table~\ref{reslntab}).    
We compute the mean offsets in $M_{20}$ ($-0.157$)and $A$ ($-0.115$) 
between the standard and high-resolution simulations after the first 0.6 Gyr.  
We recompute the standard-resolution merger time-scales ignoring the pre-merger initial galaxy morphologies and 
correcting the morphologies by the mean offsets at t $>$ 0.6 Gyr.  
These corrected time-scales are within 1$\sigma$ of the high resolution simulation merger time-scales,
where 1$\sigma$ is the standard deviation derived from the 11 viewing angles (Table~\ref{reslntab}).  
Given the large $S$ offsets with simulation resolution, we do not include $S$ in our analysis. 

\begin{figure*} 
\includegraphics[width=168mm]{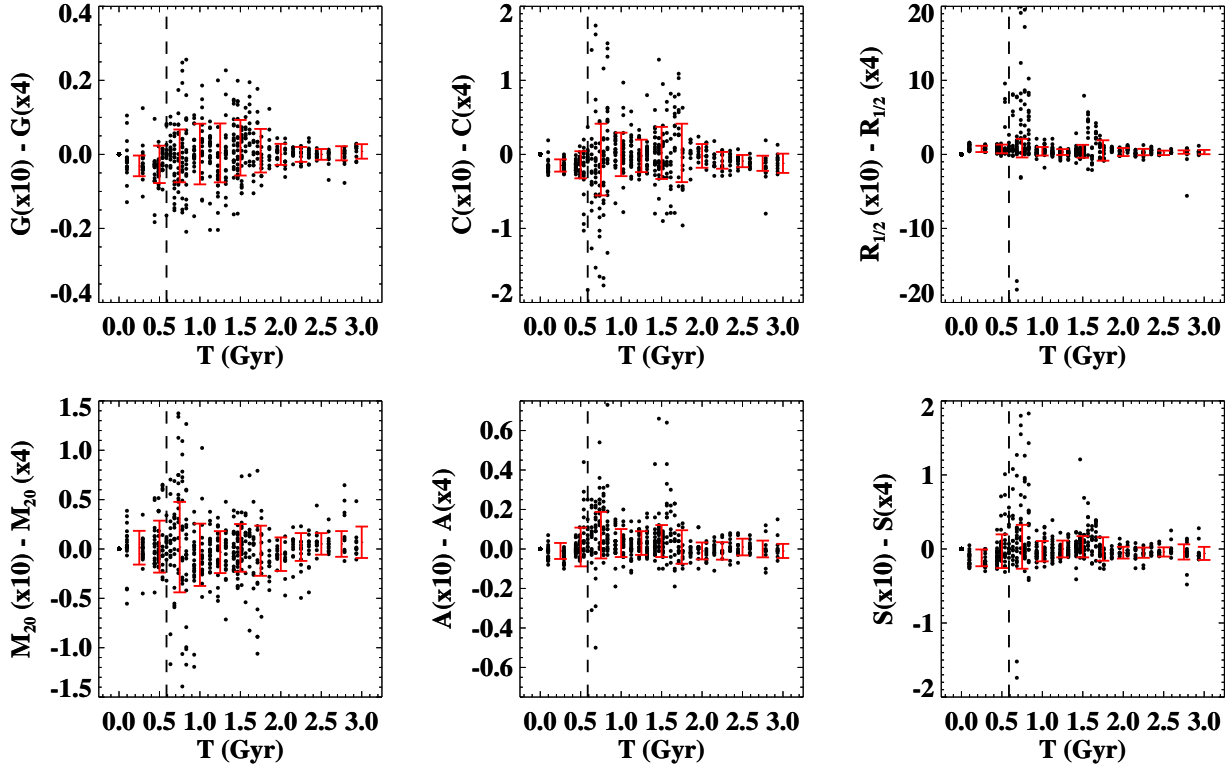}
\caption{$\Delta$ Morphology v. Time for the prograde-prograde Sbc simulation run with 
10$\times$ (SbcPPx10) and 4$\times$ (SbcPPx4) the standard number of particles.
The red error-bars show the standard deviation of the morphology differences within 0.25 Gyr bins.
The morphologies of the initial galaxies agree well at all times, including the initial stages at t $<$ 0.6 Gyr (dashed line). }\label{dif4x}
\end{figure*}

We compare the output morphologies of the 4x and 10x resolution simulations (SbcPPx4, SbcPPx10)
to check that these simulations are resolved.  Here the morphologies and time-scales agree quite well, 
even during pre-merger stage (Figure~\ref{dif4x}, Table~\ref{reslntab}).  
This work is a first attempt at computing the wavelength-dependent morphologies for a
large parameter space upon which future studies can build. Therefore we choose to examine only timesteps after the 
pre-merger stage ($t > 0.6$ Gyr) and apply the same corrections for $M_{20}$ and $A$ when calculating the 
observability time-scales for the standard-resolution simulations throughout this paper.

\begin{table*}
  \begin{minipage}{168mm}
    \caption{High Resolution Prograde-Prograde Sbc Merger (SbcPPx10) Timescales v. Viewing Angle}
    \begin{tabular}{cccccccc}
      \hline
      Cam  & T($G-M_{20}$) & T($G-A$)  & T($A$)  & T($5 < R_{p} < 20 $) & T($10 < R_{p} < 30$)      & T($10 < R_{p} < 50 $) & T($10 < R_{p} < 100 $) \\
      &   (Gyr)  & (Gyr)  & (Gyr) & (Gyr)  & (Gyr)  & (Gyr)  & (Gyr) \\
      \hline
      0 &  0.24 &  0.88 &  0.98  & 0.05 & 0.24 & 0.93 & 1.30  \\
      1 &  0.32 &  0.86 &  0.86  & 0.00 & 0.22 & 1.05 & 1.20  \\
      2 &  0.44 &  1.15 &  1.15  & 0.49 & 0.29 & 0.46 & 0.61  \\
      3 &  0.27 &  0.76 &  0.98  & 0.00 & 0.10 & 0.68 & 1.20  \\
      4 &  0.10 &  0.68 &  0.71  & 0.27 & 0.73 & 1.03 & 1.03  \\
      5 &  0.39 &  0.95 &  0.86  & 0.17 & 0.32 & 0.88 & 1.20  \\
      6 &  0.24 &  1.15 &  1.05  & 0.10 & 0.24 & 0.93 & 1.34  \\
      7 &  0.27 &  0.83 &  1.10  & 0.15 & 0.83 & 0.93 & 1.25  \\
      8 &  0.15 &  0.86 &  0.81  & 0.05 & 0.29 & 1.08 & 1.25  \\
      9 &  0.20 &  0.86 &  0.86  & 0.10 & 0.36 & 1.10 & 1.25  \\
      10 & 0.20 &  0.90 &  1.00  & 0.00 & 0.22 & 0.88 & 1.05  \\
      \hline
      mean &   0.26$\pm$ 0.10 &  0.90$\pm$ 0.14 &  0.94$\pm$ 0.13  & 0.15 $\pm$ 0.18   & 0.39 $\pm$ 0.22   & 1.00 $\pm$ 0.17   & 1.22 $\pm$ 0.20  \\
      \hline
      \end{tabular}\label{camtab}
      \medskip \\
      Computed for timesteps $t > 0.6$ Gyr, as discussed in \S 4.1. The camera angles are given in \S2.2.
      \end{minipage}
\end{table*}

\subsection{Image Resolution}
At $z>0.25$, even galaxies observed with $HST$ have images with worse spatial resolution than our fiducial
resolution (400 pc per resolution element).    As we discuss in \S3 and show in LPM04, morphologies measured in images with
spatial resolutions worse than 500 pc per resolution element have resolution-dependent biases.  
In principle,  for images with low spatial resolution one should model the redshift and PSF-dependent biases 
for one's particular dataset.    However, the turnover in the angular-size/ redshift relation is 
such that the angular scale of galaxies does not change dramatically at $z>0.6$.
In order to determine if the time-scales calculated here can be applied to $HST$ images of
high-redshift galaxies,  we measure the morphologies and time-scales of the high numerical resolution simulation SbcPPx10 tuned to 
match typical $HST$ observations of galaxies at $z \sim 1$.   
We use the TinyTim\footnote{J. Krist \& R. Hook; http://www.stsci.edu/hst/software/tinytim} 
software to calculate the PSF of $ACS$ $WFC$ in the $F814W$ (wide I) filter. 
We scale the PSF FWHM (0.14\arcsec) to 1.1 kpc to match the angular scale at $z \sim 1$ and convolve the SbcPPx10 simulation images
with this PSF.  

The morphologies measured from these images show non-negligible offsets from the images convolved
to 400 pc resolution for $M_{20}$, $C$, and $A$. The median offsets calculated for all timesteps and viewing angles 
are $\delta G = -0.01 \pm 0.06$,  $\delta M_{20} = -0.03 \pm 0.2$,  $\delta C = +0.12 \pm 0.34$ and $\delta A = -0.05 \pm 0.06$. 
These offsets are consistent with the artificial redshift tests of real galaxies by LPM04  and Conselice et al. (2005).        
Without any corrections to the merger criteria given in Eqns. 10-12, these shifts result in shorter observability time-scales  
(T($G-M_{20}$) = 0.14 $\pm$ 0.12 Gyr,  T($G-A$) = 0.58 $\pm$ 0.17 Gyr, T($A$) = 0.47 $\pm$ 0.17 Gyr vs.  0.26 $\pm$ 0.10,  0.90 $\pm$ 0.14, 
and 0.94 $\pm$ 0.13 Gyr,  respectively).  

If revised merger criteria of Eqns. 13-15 (which take into account of the effect of decreased 
spatial resolution) are applied,  then the derived time-scales are in better agreement with the higher
spatial resolution images: T($G-M_{20}$) = 0.25 $\pm$ 0.12 Gyr,  T($G-A$) = 0.70 $\pm$ 0.17 Gyr, and T($A$) = 0.58 $\pm$ 0.15 Gyr.   
The pair time-scales are also $\sim$ 200 Myr shorter than the fiducial resolution images.  Although merging objects may be more 
difficult to detect at high redshift with $HST$ observations,  the $\sim$ 200-300 Myr offsets in the asymmetry and pair time-scales are
significantly less than the variations associated with the gas properties and merger orbital properties (\S4.3 and 4.4).  
We conclude that the time-scales presented here can be applied to $HST$ observations of high-redshift 
galaxies without introducing uncertainties larger than those from the unknown distribution of merger
properties.     

\subsection{Dust and Viewing Angle}

\begin{figure*} 
\includegraphics[width=168mm]{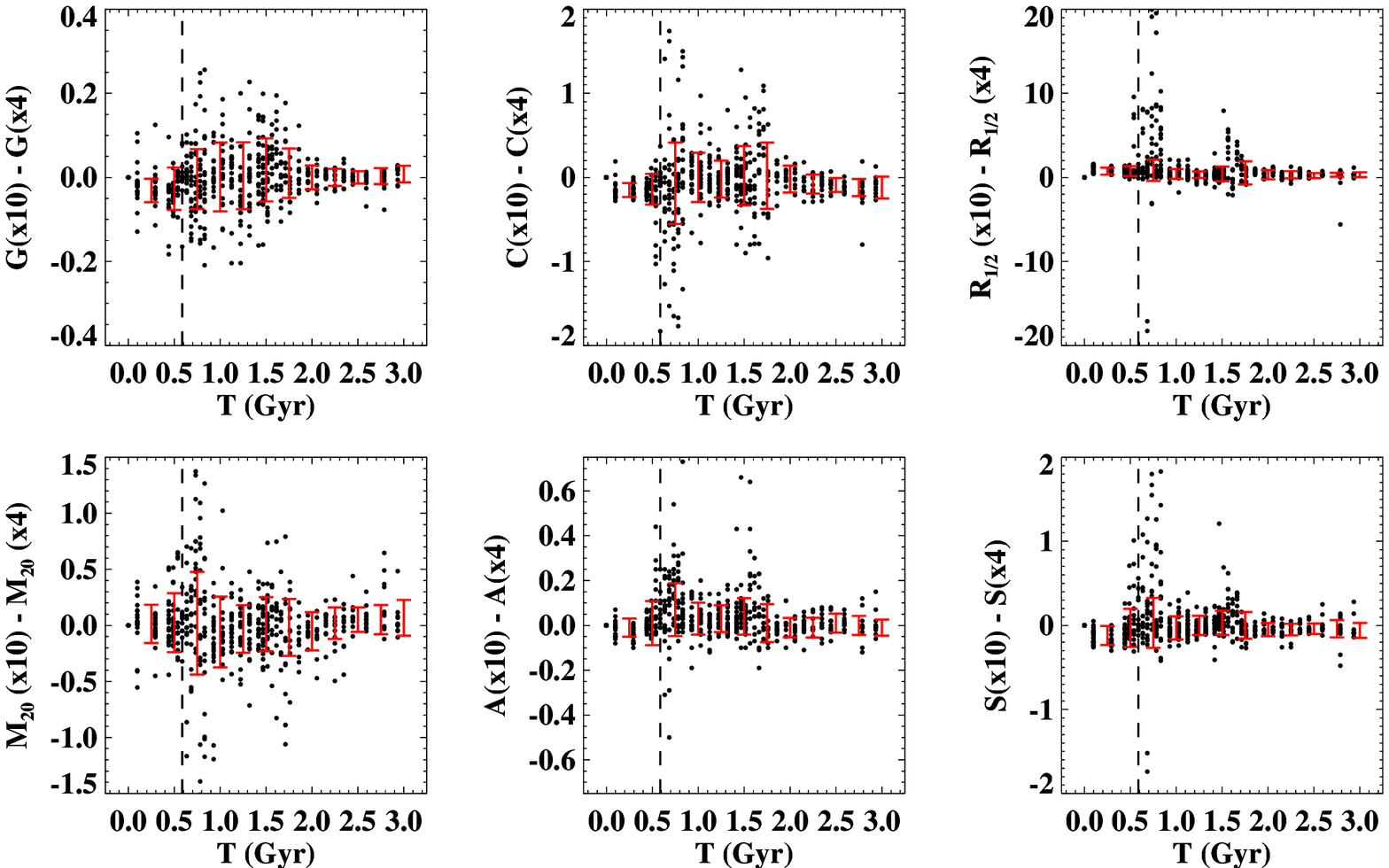}
\caption{$\Delta$ Morphology v. Time for the high numerical resolution prograde-prograde Sbc merger (SbcPPx10) with and 
without the effects of dust extinction.}\label{dust}
\end{figure*}

\begin{figure} 
\includegraphics[width=84mm]{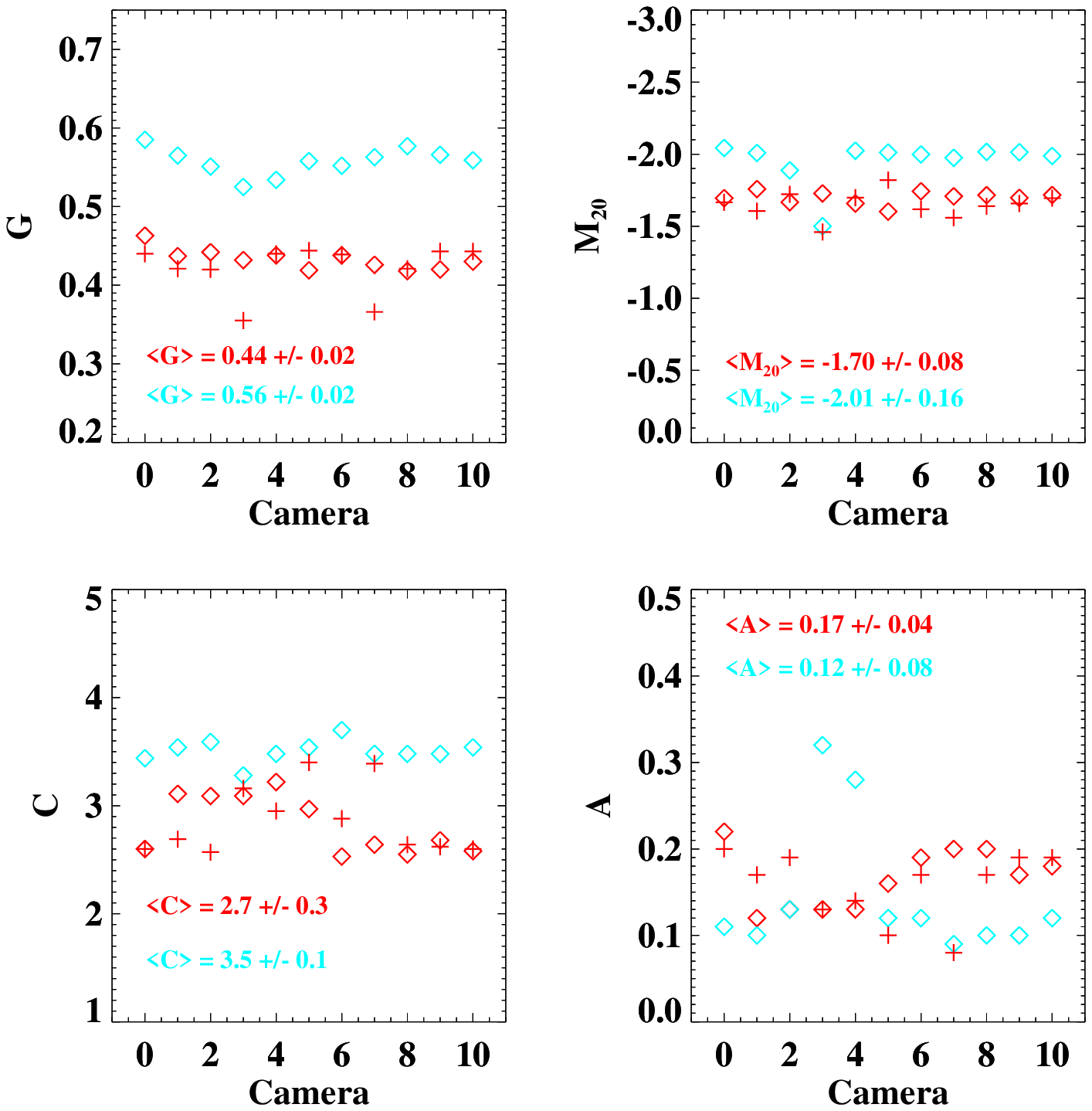}
\caption{Morphology as a function of viewing angle for the high resolution prograde-prograde Sbc merger (SbcPPx10) . The red symbols show
the initial galaxy morphologies and the cyan symbols show the final remnant morphologies. The measured 
morphologies of the initial disc galaxies do not change significantly with viewing angle.  The merger
remnant shows significantly higher $M_{20}$ and $A$ values when the dust lane is viewed
edge-on (cameras 3 and 4). }\label{inc}
\end{figure}

The presence and distribution of dust has a strong effect on the measured morphologies starting at the first pass until
after the final merger.  In Figure~\ref{dust}, we plot the difference in the morphologies when dust is and is not
included for SbcPPx10.   
During the merger, dust mitigates the effect of star-formation on the morphologies.  The
presence of dust lowers $G$ and $C$ and increases $M_{20}$ because the brightest star-forming regions near
the centres of the merging galaxies are enshrouded.  The $G-M_{20}$ observability time-scale 
is most strongly affected by the presence of dust, and is a factor of 2 less when dust is included (Table \ref{timetab}).   
The $G-A$ and $A$ time-scales are relatively independent of extinction because the measured
asymmetry is less affected (Figure~\ref{dust}).  The close pair time-scales are unchanged, as
the measured positions and projected separations are unaffected by dust extinction.
Most of the gas-rich simulations presented 
here continue to form stars at a significant rate ($>$ 2 M$_{\odot}$~yr$^{-1}$) at 1 Gyr after the
merger of the nuclei, and retain significant amounts of gas and dust. If the dust is ignored, 
the remnants appear  highly concentrated and relatively blue (Table \ref{remtab}). With dust, 
the remnants appear less concentrated with lower $C$, $G$, and higher $M_{20}$ and $A$ values 
because the central star-formation is obscured and dust increases the asymmetry of the merger 
remnant (Table \ref{remtab}).

The morphologies observed in SDSS $g$-band depend on the viewing angle, in part because the 
dust lanes will preferentially absorb blue light along certain lines of sight.  Projection effects
and the relative orientations of the merging galaxies will also change the projected separations and 
observed morphology. The scatter in the morphology at a given timestep in Figure \ref{timeSbc201a10x} is 
the result of the different viewing angles for the 11 different SUNRISE cameras.  
This scatter is largest immediately after the first pass and during the final merger when the system is most asymmetric, and
is smallest for the merger remnant which is more spherically symmetric.   We show the dependence of the measured morphologies on
viewing angle in Figure~\ref{inc} for the initial disc galaxies and the final remnant. The
measured morphologies do not change significantly with viewing angle for the initial galaxies.  
The remnant shows significantly higher $M_{20}$, $A$, and $S$ values when the final dust lane and star-forming disc is
viewed edge-on (cameras 3 and 4). 
In Table \ref{camtab}, we give the SbcPPx10 simulation 
merger time-scales for each viewing angle including the effect of dust.   
The standard deviation over all 11 viewing angles is $\sim$ 100
Myr for $T(G-M_{20})$, $T(G-A)$, and $T(A)$.   The close pair time-scales and projected separations $R_{proj}$
also depend on viewing angle, as the galaxies will have smaller $R_{proj}$ along some lines of sight.   
The close pair time-scales have a standard deviation $\sim 200$ Myr over
the 11 different viewing angles.

\subsection{Orientation and orbital parameters}


\begin{table}
    \caption{Equal-Mass Merger Morphological Timescales}
    \begin{tabular}{@{}lcccc@{}}
      \hline
      Simulation & T($G-M_{20}$)  & T($G-A$) & T($A$) \\
      & (Gyr)  & (Gyr)  & (Gyr) \\
      \hline
      \multicolumn{4}{c}{Sbc-Sbc mergers} \\
      \hline
      SbcPPx10 (no dust) &  0.44$\pm$ 0.15 &  0.89$\pm$ 0.34 &  1.12$\pm$ 0.40   \\
      \hline
      SbcPPx10       &  0.26$\pm$ 0.10 &  0.90$\pm$ 0.14 &  0.94$\pm$ 0.13   \\
      SbcPPx4        &  0.30$\pm$ 0.11 &  0.78$\pm$ 0.17 &  0.80$\pm$ 0.15   \\
      \hline
      SbcPP          &  0.39$\pm$ 0.16 &  0.78$\pm$ 0.21 &  0.74$\pm$ 0.17   \\
      SbcPR          &  0.31$\pm$ 0.10 &  0.98$\pm$ 0.11 &  1.12$\pm$ 0.11   \\
      SbcRR          &  0.60$\pm$ 0.18 &  1.33$\pm$ 0.32 &  1.46$\pm$ 0.31   \\
      SbcPol          &  0.40$\pm$ 0.25 &  1.10$\pm$ 0.24 &  1.10$\pm$ 0.29   \\
      SbcPPr-        &  0.57$\pm$ 0.20 &  0.77$\pm$ 0.19 &  0.70$\pm$ 0.16   \\
      SbcPPr+        &  1.03$\pm$ 0.74 &  0.93$\pm$ 0.48 &  1.19$\pm$ 0.57   \\
      SbcR           &  0.44$\pm$ 0.21 &  0.93$\pm$ 0.30 &  0.61$\pm$ 0.12   \\
      \hline
      SbcPPn=0       &  0.42$\pm$ 0.20 &  0.90$\pm$ 0.35 &  0.76$\pm$ 0.17   \\
      SbcRn=0        &  0.23$\pm$ 0.09 &  0.86$\pm$ 0.35 &  0.42$\pm$ 0.09   \\
      \hline
      \multicolumn{4}{c}{G-G mergers}  \\
      \hline
      G3PP           &  0.17$\pm$ 0.06 &  0.32$\pm$ 0.07 &  0.22$\pm$ 0.11   \\
      G2PP           &  0.22$\pm$ 0.14 &  0.31$\pm$ 0.15 &  0.25$\pm$ 0.19   \\
      G1PP           &  0.24$\pm$ 0.05 &  0.35$\pm$ 0.14 &  0.30$\pm$ 0.16   \\
      G0PP           &  0.30$\pm$ 0.09 &  0.43$\pm$ 0.17 &  0.39$\pm$ 0.16   \\
      \hline
      G3PPn=0        &  0.19$\pm$ 0.08 &  0.38$\pm$ 0.15 &  0.19$\pm$ 0.16   \\
      G2PPn=0        &  0.22$\pm$ 0.06 &  0.30$\pm$ 0.12 &  0.21$\pm$ 0.10   \\
      G1PPn=0        &  0.61$\pm$ 0.24 &  0.45$\pm$ 0.13 &  0.30$\pm$ 0.12   \\
      G0PPn=0        &  0.31$\pm$ 0.09 &  0.48$\pm$ 0.14 &  0.41$\pm$ 0.17   \\
      \hline
      \end{tabular}\label{timetab}
      \medskip \\
      Computed for timesteps $t > 0.6$ Gyr, as discussed in \S 4.1. 
      The standard resolution simulations also have $\delta M_{20} = -0.157$, \\ 
      $\delta A = -0.115$ correction applied, as in the last row of Table \ref{reslntab}.
\end{table}

\begin{figure*} 
\includegraphics[width=168mm]{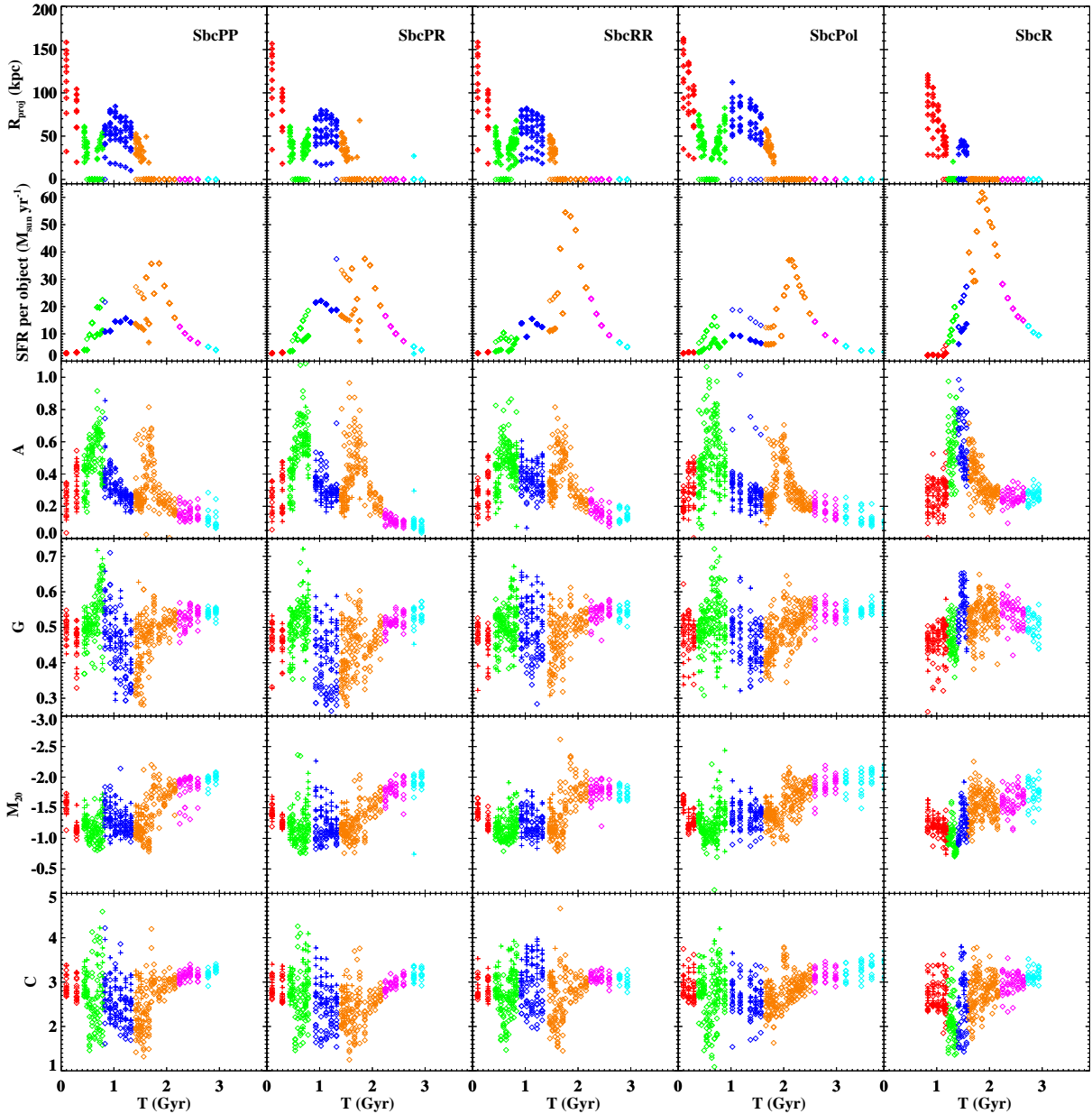}
\caption{Time v. R$_{proj}$, star-formation rate per object, $A$, $G$, $M_{20}$, and $C$ for the standard resolution prograde-prograde Sbc merger
(SbcPP), the prograde-retrograde Sbc merger (SbcPR), the retrograde-retrograde Sbc merger (SbcRR), the polar orientation Sbc merger (SbcPol), 
and the radial orbit Sbc merger (SbcR).  
Each merger stage is colour-coded as in Figure \ref{timeSbc201a10x}.  The parabolic orbit Sbc mergers show similar peaks in star-formation and
morphological disturbances at the first pass and final merger,  while the radial orbit Sbc merger (SbcR) has a single peak in both star-formation
and disturbed morphology after the initial encounter. } \label{time_orient}
\end{figure*}

\begin{figure*} 
\includegraphics[width=168mm]{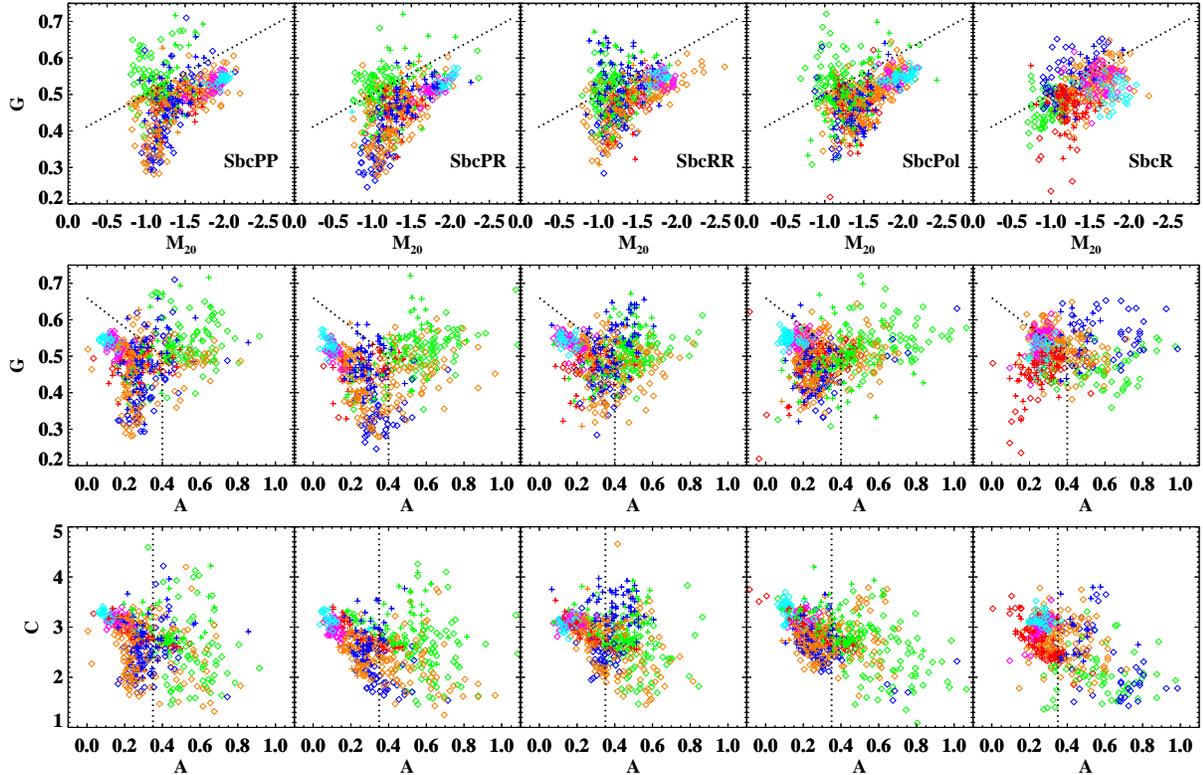}
\caption{$G-M_{20}$, $G-A$, and $C-A$ for the same simulations as Figure \ref{time_orient} (SbcPP, SbcPR, SbcRR, SbcPol, and SbcR).  
Each merger stage is colour-coded as in Figure \ref{timeSbc201a10x}. The SbcRR and SbcR simulations are more likely to have
disturbed morphologies during the maximal separation stage between the first pass and final merger (blue points) than the other Sbc simulations.
} \label{merg_orient}
\end{figure*}

\begin{figure*} 
\includegraphics[width=168mm]{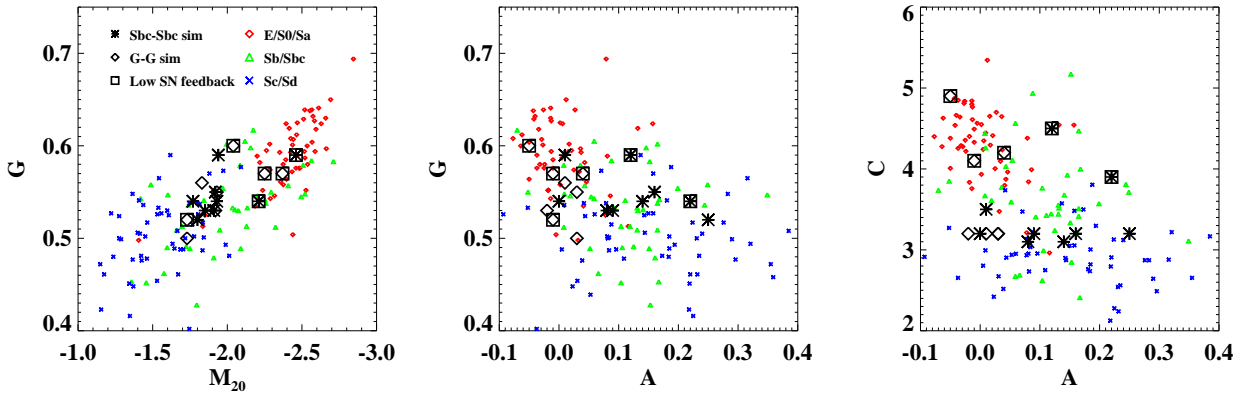}
\caption{Simulated merger remnant morphologies, including the effects of dust, for 
$G-M_{20}$, $G-A$, and $C-A$.  The Sbc-Sbc remnants are black
asterisks, and the G-G remnants are black diamonds.  Simulations run with $n=0$ supernova feedback are
surrounded by a black square.   Also plotted are the SDSS $g$ or $B$ morphologies of local galaxies from
SDSS and the Frei et al. catalog, measured by LPM04.  The remnants of all the simulations run with stiff $n=2$
supernova feedback have quantitative morphologies similar to Sb galaxies (green triangles). Some simulations with 
isothermal $n=0$ supernova feedback approach E/S0 morphologies (red diamonds). } \label{rems}
\end{figure*}

\begin{figure} 
\includegraphics[width=84mm]{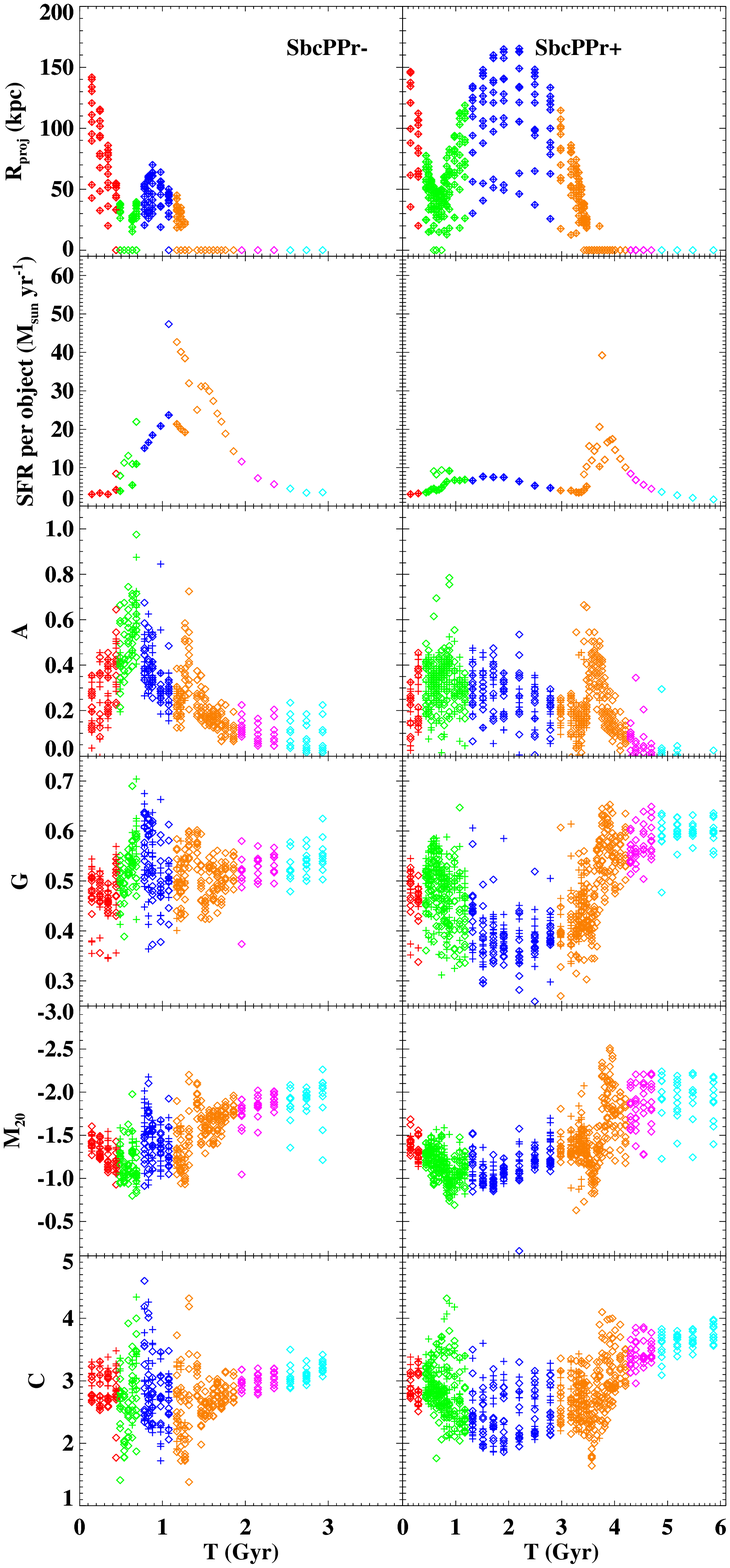}
\caption{Time v. R$_{proj}$, star-formation rate per object, $A$, $G$, $M_{20}$, and $C$ for the standard resolution prograde-prograde Sbc
mergers with small R$_{peri}$ (SbcPPr-)  and large R$_{peri}$ (SbcPPr+). 
Each merger stage is colour-coded as in Figure \ref{timeSbc201a10x}.  The large R$_{peri}$ simulation takes significantly longer to merge, and
has lower peak star-formation rates and asymmetries. } \label{time_rperi}
\end{figure}

We examine the importance of the orbits and relative orientations of the merging galaxies to their
morphologies.  Sbc mergers initialized on parabolic orbits with pericentric distances $R_{peri} = 11$ kpc 
and stiff feedback were simulated with roughly prograde-prograde (SbcPP), 
prograde-retrograde (SbcPR), retrograde-retrograde (SbcRR), and polar (SbcPol) orientations. 
All of these simulations have similar orbital decay times, with the polar orientation merger taking a few 100 Myr longer
for the final merger to occur (Table 3).   All of the parabolic Sbc simulations experience peaks 
in star-formation and asymmetry at the first pass and the final merger, with the maximum star-formation rate 
depending on the relative orientation of the discs (Figure \ref{time_orient}). 
The strength of the morphological disturbances also depends 
on the orientation of the galaxies, with the intrinsically asymmetric polar and
retrograde-prograde mergers showing the highest asymmetries (Fig. \ref{time_orient}).   The viewing-angle averaged time-scales 
during which a particular set of quantitative morphologies are disturbed vary by a factor of 2 for these different orientations
($T(G-M_{20})$ $\sim$ 0.3-0.6 Gyr, $T(G-A) \sim$ 0.8-1.3 Gyr, and $T(A) \sim$ 0.7-1.5 Gyr; Table \ref{timetab}, 
Figure \ref{merg_orient}). This is in reasonable agreement with the $\sim$ 0.7-1.0 Gyr asymmetry time-scales 
of the star-particles of equal-mass prograde-inclined, retrograde-inclined, and prograde-retrograde merger 
simulations found by Conselice (2006).  The retrograde-retrograde merger is disturbed for the longest time 
for all of the quantitative morphology measures.    The time-scales are also sensitive to the criteria used to identify the merger 
(Table \ref{timetab}; Figure \ref{merg_orient}).  The typical
$T(G-M_{20})$ is $\sim$ 0.4 Gyr, while $T(A)$ $\sim$ 1.1 Gyr and $T(G-A) \sim$ 1.0 Gyr.
The close pair time-scales do not vary strongly with orientation ($\delta T \sim$ 200 Myr; Table \ref{timetab2})  
but do depend on the range of $R_{proj}$ chosen,  with $R_{proj} < 30$ kpc time-scales often significantly  
shorter than typical $T(G-A)$ and $T(A)$ values.  All of the simulations show enhanced star-formation for significantly
longer periods than the morphological disturbances, with the peaks in the star-formation rates often occurring after
the peaks in the asymmetry.   
The remnant morphologies are generally similar (Table \ref{remtab}), although the retrograde-retrograde 
merger remnant has a higher $M_{20}$ value than the prograde-prograde merger ($-1.77$ v. $-1.93$) and both 
the retrograde-retrograde and polar merger remnants 
have higher asymmetries (0.15 v. 0.0).  The remnant morphologies are more consistent with
early-type disc galaxies (Sb) than spheroids (E/S0) (Figure \ref{rems}). The remnants are forming stars at 
$\sim 5-6$ M$_{\sun}$ yr$^{-1}$.

A sub-parabolic Sbc-Sbc merger simulation with a highly radial orbit, zero net angular momentum, prograde-retrograde 
orientation and stiff feedback was also examined (SbcR).  Because the galaxies start with low relative velocities, 
it takes significantly longer for the first pass to occur (1.3 Gyr v. 0.6 Gyr).  However, there is 
significantly less time between the first pass and the final merger
(0.3 Gyr v. 1.1 Gyr; Table \ref{timetab}).  As a result, the morphologies and star-formation rates as a function of merger
stage are quite different from the parabolic orbit simulations (Fig. \ref{time_orient}, \ref{merg_orient}).  
Asymmetry peaks during the first pass and maximal separation stages, but is quite low during the final merger.  
On the other hand, the star-formation rate is strongly enhanced throughout the later merger stages 
and reaches its peak during the final merger.  The $G-M_{20}$, $G-A$, and $A$ time-scales are  similar to 
the parabolic $R_{peri}=11$ Sbc simulations.  The close pair time-scales, however,  are naturally $\sim$ 50\% shorter than
parabolic orbits. The merger remnant has a large bulge surrounding by a very blue star-forming ring.  It has the
highest star-formation rate (13 M$_{\odot}$ yr$^{-1}$) of any of the simulations, and because of the bright blue ring, 
its morphology is the most asymmetric (0.25) and disc-like in its $G$, $M_{20}$ and $C$ values. 

Two additional prograde-prograde parabolic orbit simulations with smaller and larger pericentric distances ($R_{peri} = 5.5$, 
44 kpc v. 11 kpc) were analysed. The small R$_{peri}$ simulation (SbcPPr-) takes 340 Myr less to merge, 
while the large R$_{peri}$ 
(SbcPPr+) simulation takes an additional 2 Gyr to merge (Table \ref{timetab}, Figure \ref{time_rperi}).  
The merger time-scales and properties of the small R$_{peri}$
simulation are similar to the fiducial SbcPP simulation, with somewhat shorter close pair time-scales
for R$_{proj}$ $<$ 50 and 100 $h^{-1}$ kpc (Table \ref{timetab2}, Figure \ref{time_rperi}). 
The large $R_{peri}$ simulation, however, has little enhanced star-formation and
lower asymmetries during the first pass, and experiences less enhanced star-formation during the final merger
because more gas has been consumed in `normal' disc star-formation (Fig. \ref{time_rperi}). 
Despite weaker morphological disturbances, the $G-M_{20}$ observability time-scales 
for the large $R_{peri}$ simulations is significantly larger (1.0 Gyr v. 0.4 Gyr) but with larger scatter with viewing
angle. The $G-A$ and $A$ time-scales are also longer with larger scatter, as are the 
the close pair time-scales for $R_{proj}$ $<$ 50 and 100 kpc.  
The large $R_{peri}$ remnant has somewhat higher $G$, $M_{20}$ and $C$ values than the 
$R_{peri}$ = 5.5 and 11 kpc remnants, making its morphology more like typical spheroidals (Table \ref{remtab}). Visual inspection
of the remnant shows that the recent star-formation in the large $R_{peri}$ is much more centralized, while
the smaller $R_{peri}$ simulations have an extended disc of young stars. Because the SbcPPr+ simulation takes twice as
long to merge, most of its remnant's cold gas has been consumed in star-formation during the merger.

In summary, we find that orientation and large pericentric distances can have a significant effect on the time-scales during
which mergers can be identified morphologically. Some relative orientations of the merging system increase the 
strength of the morphological disturbances (prograde-retrograde, polar), while other orientations increase the time-scales 
of those disturbances (retrograde-retrograde).
Large pericentric distances naturally result in long orbital decay times, which suggests that
the duration of the merger is as important to the time-scales of morphological disturbances as
the orientation of the merging galaxies. The orbits do affect the timing of the morphological disturbances. 
Most of the parabolic-orbit simulations show peaks in morphological disturbances at the first pass and final merger. 
The highly radial orbit simulation shows a single peak in asymmetry during the first pass, 
while the large pericentric radius simulation shows a less
dramatic enhancement of asymmetry during the first pass and final merger.  However, the highly radial orbit 
simulation also has morphological disturbance time-scales that agree with the parabolic orbits to within the scatter
associated with the viewing angle.
Enhanced star-formation rates generally occur for longer durations than the morphological disturbances, with the
star-formation rates often peaking after the asymmetries. 
The remnants all have similar concentrations ($G$, $M_{20}$, and $C$) consistent with early-type spirals 
but only the retrograde-retrograde, polar, and highly radial orbit merger remnants show significant asymmetries ($A > 0.1$). 

\begin{figure} 
\includegraphics[width=84mm]{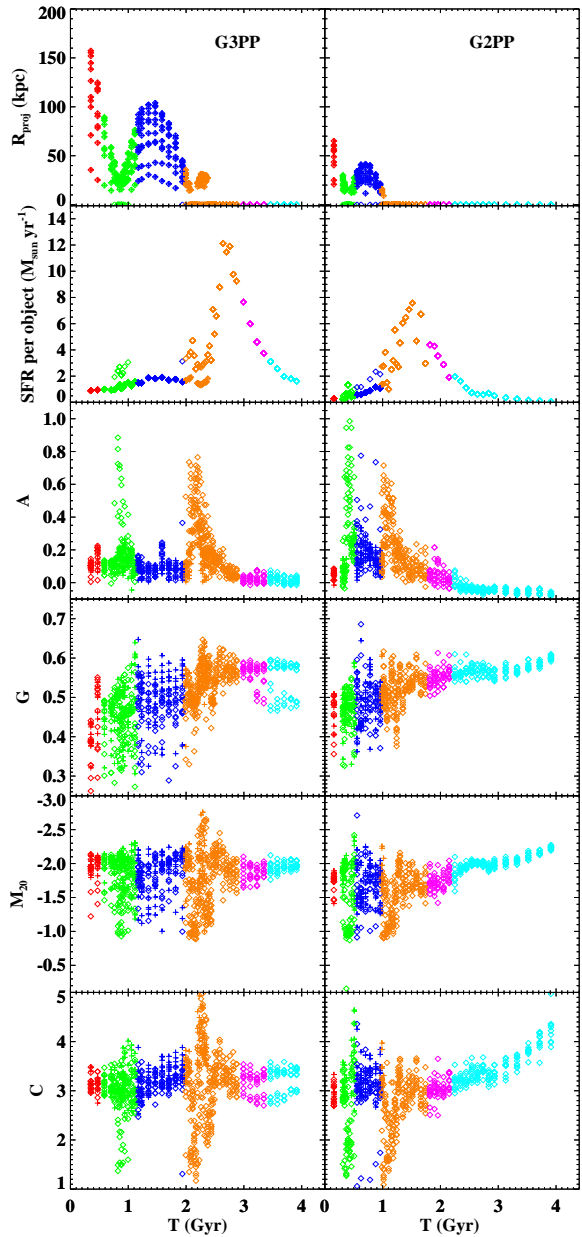}
\caption{Time v. R$_{proj}$, star-formation rate per object, $A$, $G$, $M_{20}$, and $C$ for the prograde-prograde 
$1.2 \times 10^{12} \Msun$ G3 merger (G3PP) and less massive prograde-prograde $5 \times 10^{11} \Msun$ G2 merger (G2PP). 
Each merger stage is colour-coded as in Figure \ref{timeSbc201a10x}. The G3 merger is less morphologically disturbed than the Sbc prograde-prograde
merger (SbcPP) during the first pass,  and experiences a peak in the star-formation rate well after the peak in asymmetry at the final merger.
} \label{time_mass1}
\end{figure}

\begin{figure} 
\includegraphics[width=84mm]{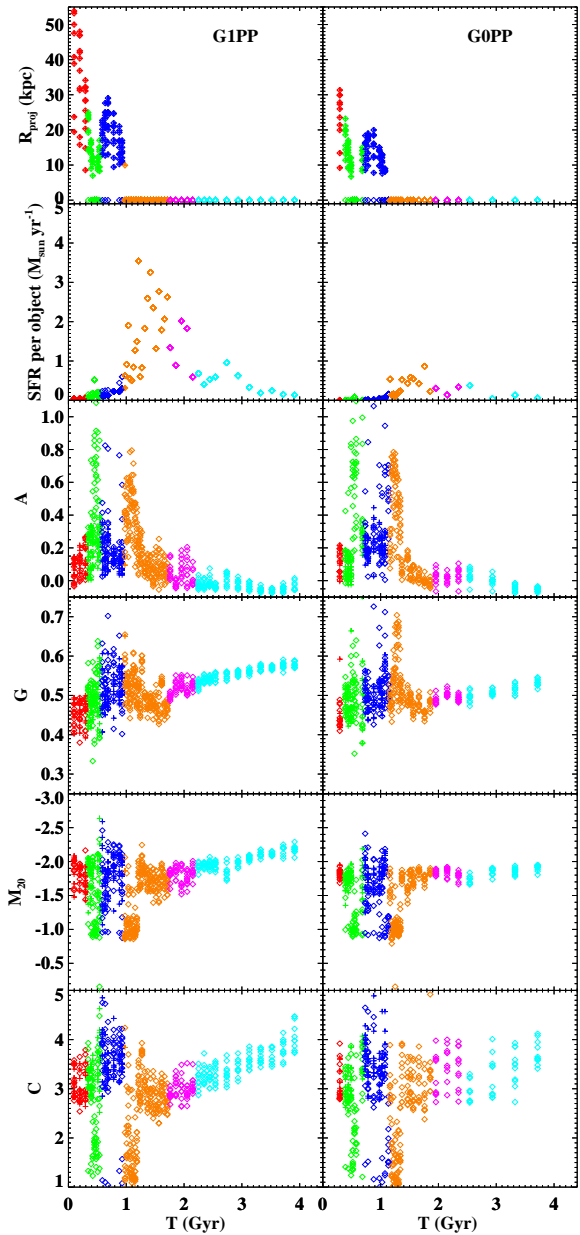}
\caption{Time v. R$_{proj}$, star-formation rate per object, $A$, $G$, $M_{20}$, and $C$ for the prograde-prograde low-mass $2 \times 10^{11} \Msun$ G1
and $5 \times 10^{10} \Msun$ G0 mergers (G1PP, G0PP).  Each merger stage is colour-coded as in Figure \ref{timeSbc201a10x}.  The lower mass mergers undergo
less star-formation but have time-scales for disturbed morphology similar to the more massive G3 and G2 mergers.} \label{time_mass2}
\end{figure}

\subsection{Gas fraction and scalelength}
Gas-rich mergers undergo significant star-bursts triggered by the varying tidal forces and
inflow of gas during the merger. Because these new stars can influence the measured morphologies, 
it is likely that the amount of gas available to form stars affects morphology time-scales 
during the merger process.  The Sbc galaxy and the G3 galaxy have similar total masses 
($8.1 \times 10^{11}$ and $1.2 \times 10^{12}$ M$_{\sun}$, respectively), 
and similar bulge to disc stellar mass ratios (0.25 and 0.21). 
However, the Sbc galaxy has a much larger gas reservoir with over 50\% of its baryons in gas.  
The G3 galaxy, on the other hand, has only $\sim$~20\% of its baryons in gas.   Both models assume that the gas
disc has a scalelength three times the scalelength of the stellar disc, but the Sbc's gas and stellar
discs are twice the adopted scalelengths for the G model.  The end result is that the 
Sbc merger simulations have much more gas at large radii as well as a higher
density of gas within the central regions.  Cox et al. (2008) found that higher central gas densities lead to 
less merger-induced star-formation when compared to the undisturbed disc star-formation.  The SbcPP merger
has less efficient merger-driven star-formation throughout the interaction relative to the G3PP merger
(where efficiency refers to the fraction of total gas converted to stars),  with 
the SbcPP and G3PP mergers showing 23\% and 46\% more star-formation than their undisturbed counterparts respectively.   However, 
the SbcPP merger experiences higher star-formation rates in general and at the first pass and final merger in particular, 
reflecting its high gas fraction. 

We compare the time-dependent morphologies and projected separations for prograde-prograde parabolic orbit mergers with stiff
feedback for the Sbc and G3 galaxies (SbcPP and G3PP; Figs. \ref{time_orient} and \ref{time_mass1}).  
We find that the merger time-scales and morphologies are also affected by the
gas disc properties.  Despite similar pericentric distances, the G3PP simulation takes about 700 Myr longer
for the nuclei to coalesce than the SbcPP simulation. 
As a result, the time-scales during which the merging galaxies can be identified as a close pair are also longer.  
The G3PP simulation spends three times longer as a very close pair ($R_{proj} < 30$  $h^{-1}$ kpc) than 
the SbcPP simulation  (Table \ref{timetab2}).  However, the
time-scales for morphological disturbances are shorter for the lower gas fraction G3PP 
simulation by a factor of $\sim$ 2-4 (Table \ref{timetab}). 
Although the G3PP merger is more efficient at turning the available gas into stars than the SbcPP merger, the G3PP merger 
has less star-formation overall because of its larger bulge and lower gas fraction (Fig. \ref{time_mass1}, \ref{merg_mass}).
The G3PP merger experiences less star-formation along tidal arms and lower asymmetries at the first pass
than the SbcPP merger. 
Nevertheless, the remnants have similar quantitative morphologies and star-formation rates (Table \ref{remtab}).  
Both simulations produce dusty remnants with $G$, $M_{20}$, $C$, and $A$ values consistent with bulge-dominated spirals (Fig. \ref{rems}).
Both remnants have significant residual star formation (3-5 M$_{\sun}$ yr$^{-1}$) and dust reddening.

\subsection{Mass}

\begin{figure*} 
\includegraphics[width=168mm]{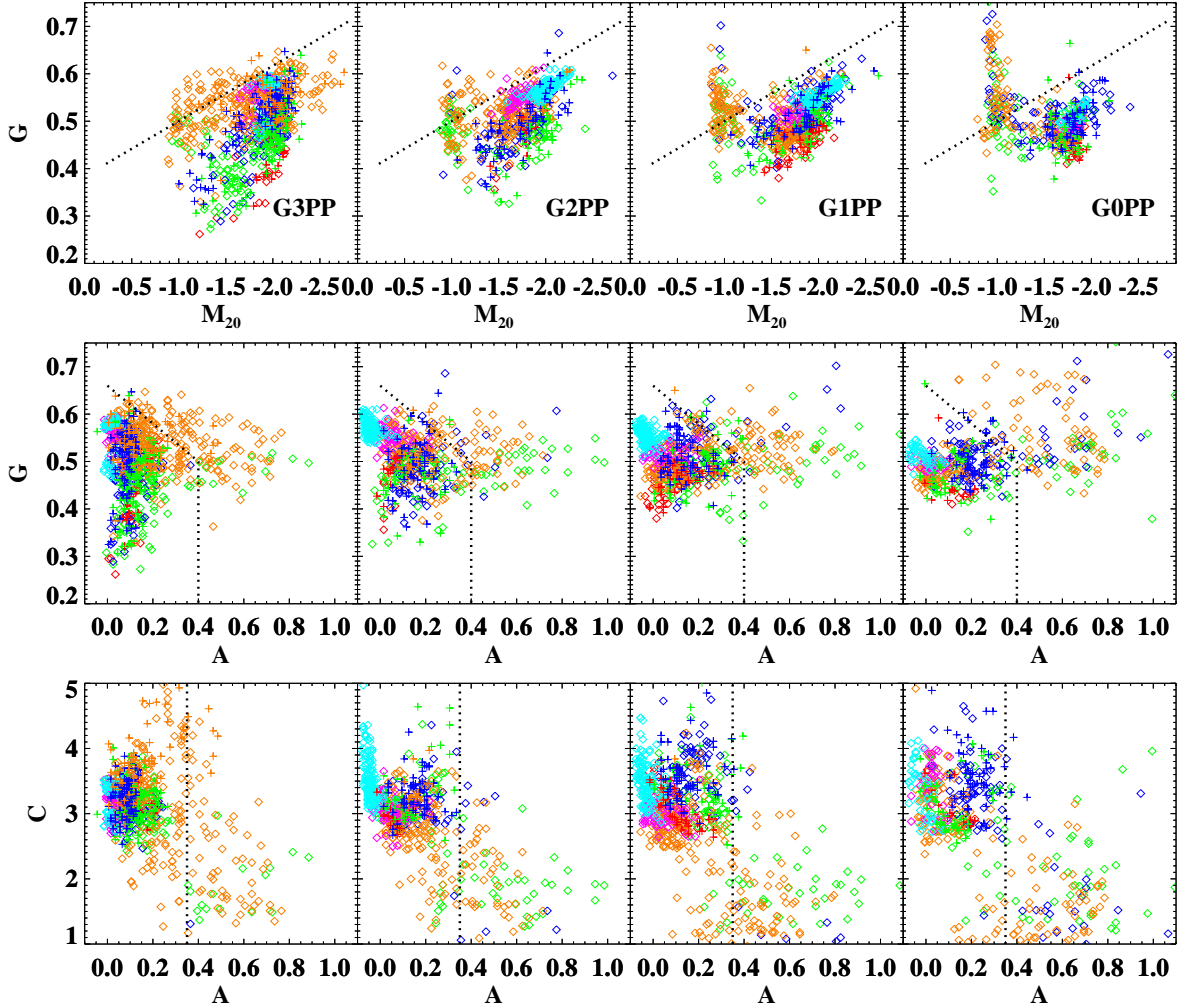}
\caption{$G-M_{20}$, $G-A$, and $C-A$ for the prograde-prograde 
G3PP, G2PP, G1PP, and G0PP simulations. The simulations span a factor of 23 in total mass, where 
the virial mass of the initial galaxy is $1.2 \times 10^{12} \Msun$ for G3, $5.1 \times 10^{11} \Msun$ for G2, 
$2.0 \times 10^{11} \Msun$ for G1,  and $5.1 \times 10^{11}$ for G0. 
Each merger stage is colour-coded as in Figure \ref{timeSbc201a10x}.  Unlike the Sbc mergers, the G-series simulations are only detected
at the final merger (orange points). } \label{merg_mass} 
\end{figure*}
The total mass involved in the merger may also affect morphologies and star-formation rates, as larger galaxies
have deeper potential wells and produce stronger tidal forces.  Equal-mass prograde-prograde merger simulations 
spanning a factor of 23 in total mass and a factor of 50 in stellar mass
were examined to explore the effects of merger mass (G3PP, G2PP, G1PP, and G0PP simulations).  
The progenitor galaxies have increasing gas fractions and total
mass to light ratios with decreasing mass (Table 1).  Both supernova feedback models were explored for all the G-series mergers. 
The orbits for all of the G-series mergers are slightly sub-parabolic, with eccentricities $e=0.95$.   
This significantly shortens the decay times for the G1 and G0 mergers, and hence may result in shorter
close pair time-scales than would be observed for $e=1.0$ orbits.   We do not expect this to impact the
morphology observability time-scales, as disturbed morphologies are apparent only at the first pass and 
final merger stages.    The initial separations are less than 100 $h^{-1}$ kpc for the G2, G1, and G0 mergers, and
so close pair time-scales are not computed when the initial separation is less than the measured 
range of projected separations (Table \ref{timetab2}).

We find that all of the equal-mass G-series mergers show similar correlations of the morphologies with merger stage. 
The morphological disturbance time-scales are $\sim$ 100 -200 Myr longer for the lowest mass merger (G0PP) than the
highest mass merger (G3PP; Table \ref{timetab}, Fig. \ref{time_mass1}, \ref{time_mass2}).  However, the time between the first pass 
and coalescence of the nuclei is $\sim$ a factor of 2 less for the lower mass merger than the highest mass merger 
(Table \ref{timetab}, Figure \ref{merg_mass}), and the close pair time-scales at $10 < R_{proj} < 30$ $h^{-1}$ kpc  
reflect this (Table \ref{timetab2}). 
The close pair time-scales at $5 < R_{proj} < 20$ $h^{-1}$ kpc show longer time-scales for the lowest mass merger. This is an artifact of the
object detection algorithm.  Larger galaxies are most likely to be counted as one object at small separations because they have larger
scalelengths and overlapping isophotes.
The merger remnants for all but the lowest mass merger (G0PP) are remarkably similar in their morphology, with the
lower mass remnants showing lower star-formation rates and extinctions.   The G0PP remnants are 
more disc-like in their $G$ and $M_{20}$ values, even when $n=0$ feedback is adopted.

\subsection{Supernova feedback}

\begin{figure*} 
\includegraphics[width=168mm]{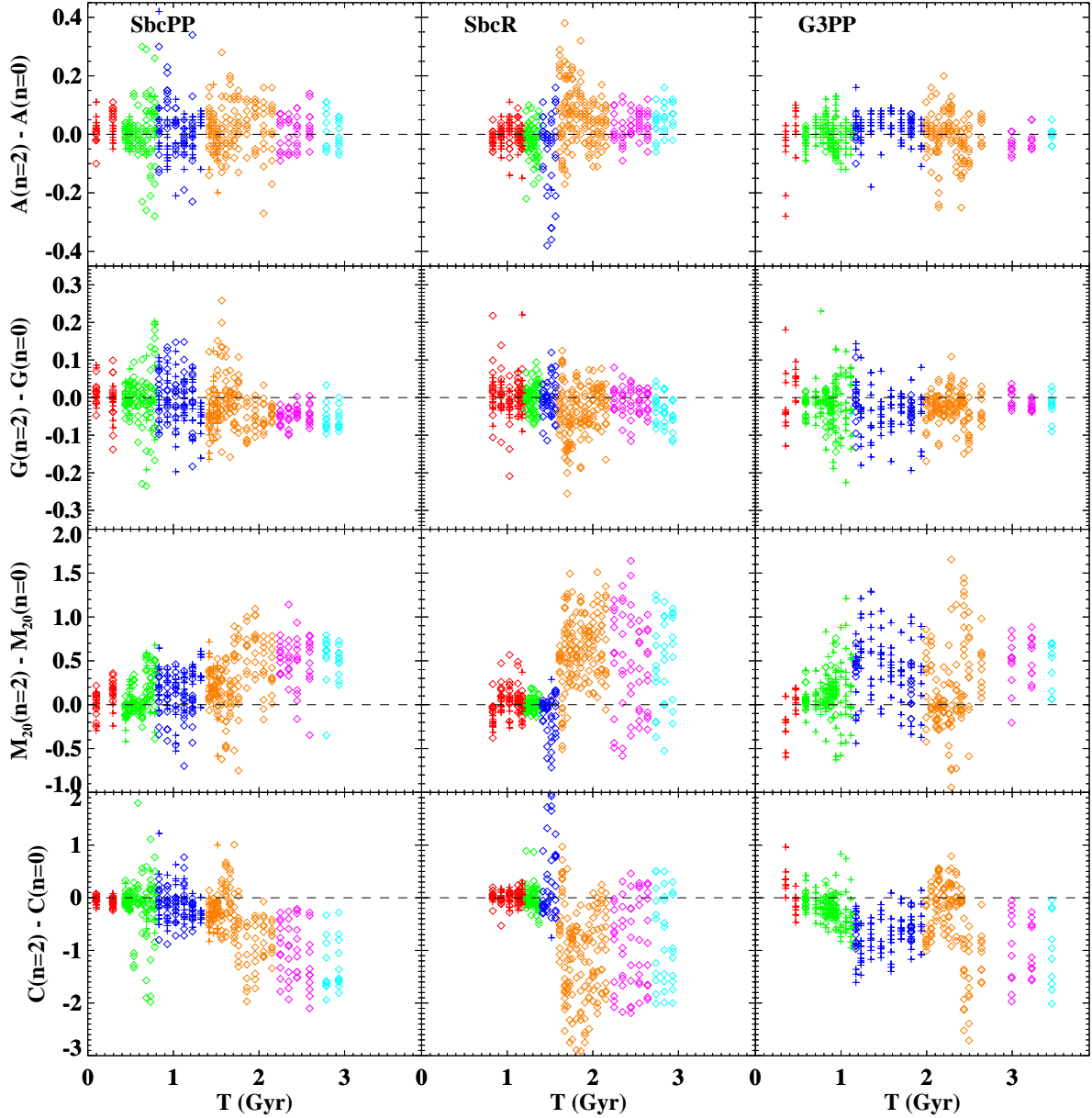}
\caption{ $\Delta$ Morphology v. time  for the simulations
with different supernovae feedback prescriptions (SbcPP = prograde-prograde Sbc; SbcR = radial orbit Sbc; G3PP = prograde-prograde G3). 
The $M_{20}$ values are higher and the concentrations are lower for $n=2$ `stiff' supernovae feedback simulations during and after
the final merger stage (orange, magenta, and cyan points). Each merger stage is colour-coded as in Figure \ref{timeSbc201a10x}.} \label{time_feedback}
\end{figure*}

The parabolic prograde-prograde and radial prograde-retrograde Sbc simulations and all of the G-series 
simulations were run with both 
supernova feedback models.  Although the total gas consumption and star formation during the merger
are similar for both feedback scenarios (Cox et al. 2006, 2008), 
the isothermal $n=0$ feedback simulations with parabolic orbits experience more star-formation 
during the first pass and have less gas available for a second starburst during the final merger (Cox et al. 2006).
The opposite is true for the highly radial orbit Sbc simulations because there is not enough time between
the first pass and final merger for the low feedback simulation to consume large amounts of gas.  
The $n=0$ radial-orbit Sbc merger experience the highest peak star-formation rate of all the merger 
simulations during the final merger ($> 500$ M$_{\odot}$ 
yr$^{-1}$), while the $n=2$ radial-orbit Sbc merger experiences its peak star-formation rate during the first pass
($\sim$ 80 M$_{\odot}$ yr$^{-1}$). 

We find that close pair time-scales and morphological disturbances as a function of merger stage and merger
time-scales are generally similar for the different feedback models, given the scatter with viewing angle 
(Tables \ref{timetab}, \ref{timetab2}).
The primary difference between the $n=2$ stiff feedback and the $n=0$ isothermal feedback
simulations appears in the properties of the remnants (Table \ref{remtab}, Fig. \ref{rems}).  
The $n=0$ feedback remnants are significantly more like E/S0 in
the quantitative morphologies ($G$, $M_{20}$ and $C$) because they have $\sim$ 40-50\% lower gas metallicities and hence
less dust to obscure the nuclei.  The lower gas metallicities are probably an artifact of our chemical enrichment scheme, rather than
a robust prediction of the dust evolution.   Supernova are assumed to produce only metals which enrich 
surrounding gas particles, but do not produce any gas particles themselves.   If all surrounding
gas particles are consumed in star-formation,  the metals which would have be produced in supernova 
have no place to go and remain locked up in the stars.  The $n=0$ feedback models experience more intense star-formation and
consume a great amount of their existing gas during the first pass,  hence are more effected by these limitations in our model. 
Nevertheless, while it is unclear if the dust properties of the merger remnants will be strongly affected by
feedback, it is likely that the $n=2$ feedback remnants have too much dust, as we do not include any dust destruction mechanisms.

\section{DISCUSSION}


\begin{table*}
  \centering
  \begin{minipage}{140mm}
    \caption{Equal-Mass Close Pair Timescales}
    \begin{tabular}{@{}lcccc@{}}
      \hline
      Simulation & T($5 < R_{proj}   < 20 $) & T($10 < R_{proj}  < 30$) & T($10 < R_{proj}  < 50 $) 
      &T($10 < R_{proj}   < 100 $) \\
      & (Gyr)  & (Gyr)  & (Gyr)  & (Gyr) \\
      \hline
      \multicolumn{5}{c}{Sbc-Sbc mergers} \\
      \hline
      SbcPPx10      & 0.15 $\pm$ 0.18   & 0.39 $\pm$ 0.22   & 1.00 $\pm$ 0.17   & 1.22 $\pm$ 0.20 \\
      SbcPPx4       & 0.26 $\pm$ 0.21   & 0.53 $\pm$ 0.31   & 1.12 $\pm$ 0.17   & 1.34 $\pm$ 0.19 \\
      \hline 
      SbcPP         & 0.15 $\pm$ 0.19   & 0.35 $\pm$ 0.23   & 0.90 $\pm$ 0.16   & 1.20 $\pm$ 0.18   \\  
      SbcPR         & 0.27 $\pm$ 0.49   & 0.54 $\pm$ 0.50   & 1.08 $\pm$ 0.59   & 1.37 $\pm$ 0.55   \\
      SbcRR         & 0.16 $\pm$ 0.26   & 0.44 $\pm$ 0.35   & 0.97 $\pm$ 0.24   & 1.31 $\pm$ 0.14   \\
      SbcPPr-       & 0.13 $\pm$ 0.10   & 0.43 $\pm$ 0.16   & 0.78 $\pm$ 0.11   & 0.99 $\pm$ 0.15   \\
      SbcPPr+       & 0.26 $\pm$ 0.29   & 0.50 $\pm$ 0.43   & 1.38 $\pm$ 0.92   & 3.06 $\pm$ 0.52   \\
      SbcPol         & 0.10 $\pm$ 0.08   & 0.34 $\pm$ 0.18   & 0.92 $\pm$ 0.38   & 1.45 $\pm$ 0.26   \\
      SbcR          & 0.08 $\pm$ 0.20   & 0.18 $\pm$ 0.18   & 0.36 $\pm$ 0.18   & 0.84 $\pm$ 0.09   \\
      \hline  
      SbcPPn=0      & 0.15 $\pm$ 0.20   & 0.43 $\pm$ 0.28   & 0.96 $\pm$ 0.17   & 1.27 $\pm$ 0.15   \\
      SbcRn=0       & 0.03 $\pm$ 0.06   & 0.13 $\pm$ 0.19   & 0.29 $\pm$ 0.21   & 0.77 $\pm$ 0.05   \\  
      \hline
      \multicolumn{5}{c}{G-G mergers}\\
      \hline
      G3PP          & 0.39 $\pm$ 0.30   & 0.72 $\pm$ 0.39   & 1.21 $\pm$ 0.38   & 1.85 $\pm$ 0.13   \\
      G2PP \footnote{Initial separation is 70 $h^{-1}$ kpc.}  & 0.43 $\pm$ 0.20   & 0.60 $\pm$ 0.16   & 0.71 $\pm$ 0.20   &  $-$             \\  
      G1PP \footnote{Initial separation is 56 $h^{-1}$ kpc.}  & 0.58 $\pm$ 0.13   & 0.52 $\pm$ 0.16   & 0.64 $\pm$ 0.20   & $-$  \\  
      G0PP \footnote{Initial separation is 42 $h^{-1}$ kpc.}   & 0.67 $\pm$ 0.19   & 0.39 $\pm$ 0.22   &  $-$    &  $-$  \\  
      \hline
      G3PPn=0       & 0.30 $\pm$ 0.38   & 0.56 $\pm$ 0.39   & 1.10 $\pm$ 0.37   & 1.72 $\pm$ 0.13   \\
      G2PPn=0$^a$       & 0.39 $\pm$ 0.16   & 0.62 $\pm$ 0.17   & 0.70 $\pm$ 0.20   & $-$ \\  
      G1PPn=0$^b$       & 0.54 $\pm$ 0.09   & 0.50 $\pm$ 0.17   & 0.54 $\pm$ 0.20   &  $-$     \\     
      G0PPn=0$^c$       & 0.64 $\pm$ 0.17   & 0.38 $\pm$ 0.21   & $-$         & $-$  \\  
      \hline
      \end{tabular} \label{timetab2}
      \medskip \\
      $R_{proj}$ has units $h^{-1}$ kpc.  Timescales for simulations with starting separations less than
      maximum $R_{proj}$ are not calculated.
     \end{minipage}
\end{table*}

\begin{table*}
  \centering
  \begin{minipage}{140mm}
    \caption{Equal-Mass Merger Remnant Properties}
    \begin{tabular}{@{}lccccc@{}}
      \hline
      Simulation & $G$  & $M_{20}$  & $C$ & $A$ & $SFR$ (M$_{\odot}$ yr$^{-1}$) \\
      \hline
      \multicolumn{6}{c}{Sbc-Sbc mergers} \\
      \hline
      SbcPPx10 (no dust) & 0.59 $\pm$ 0.03  & -1.75 $\pm$ 0.39   & 4.1 $\pm$ 0.3  & 0.05 $\pm$ 0.02  & 5.82 \\
      \hline
      SbcPPx10          & 0.54 $\pm$ 0.01  & -1.80 $\pm$ 0.23   & 3.4 $\pm$ 0.3  & 0.15 $\pm$ 0.07  & 5.82 \\
      SbcPPx4           & 0.55 $\pm$ 0.01  & -1.98 $\pm$ 0.07   & 3.6 $\pm$ 0.1  & 0.16 $\pm$ 0.05  & 5.28 \\
      \hline
      SbcPP             & 0.54 $\pm$ 0.01  & -1.93 $\pm$ 0.09   & 3.2 $\pm$ 0.1  & 0.00 $\pm$ 0.48  & 5.24 \\
      SbcPR             & 0.53 $\pm$ 0.02  & -1.91 $\pm$ 0.11   & 3.2 $\pm$ 0.1  & 0.09 $\pm$ 0.02  & 5.11 \\
      SbcRR             & 0.54 $\pm$ 0.02  & -1.77 $\pm$ 0.11   & 3.1 $\pm$ 0.1  & 0.14 $\pm$ 0.04  & 6.78 \\
      SbcPPr-             & 0.53 $\pm$ 0.03  & -1.85 $\pm$ 0.20   & 3.1 $\pm$ 0.2  & 0.08 $\pm$ 0.10  & 4.58 \\
      SbcPPr+             & 0.59 $\pm$ 0.04  & -1.94 $\pm$ 0.28   & 3.5 $\pm$ 0.2  & 0.01 $\pm$ 0.10  & 3.68 \\
      SbcPol             & 0.55 $\pm$ 0.02  & -1.92 $\pm$ 0.14   & 3.2 $\pm$ 0.2  & 0.16 $\pm$ 0.05  & 5.45 \\
      SbcR              & 0.52 $\pm$ 0.03  & -1.80 $\pm$ 0.22   & 3.2 $\pm$ 0.2  & 0.25 $\pm$ 0.04  & 12.85\\
      \hline
      SbcPPn=0            & 0.59 $\pm$ 0.04  & -2.46 $\pm$ 0.38  & 4.5 $\pm$ 0.6  & 0.12 $\pm$ 0.08  & 3.41 \\
      SbcRn=0             & 0.54 $\pm$ 0.04  & -2.21 $\pm$ 0.49  & 3.9 $\pm$ 0.9  & 0.22 $\pm$ 0.04  & 8.33 \\
      \hline
      \multicolumn{6}{c}{G-G mergers} \\
      \hline
      G3PP                & 0.55 $\pm$ 0.05 & -1.93 $\pm$ 0.10  & 3.2 $\pm$ 0.3  & 0.03 $\pm$ 0.02  & 3.11 \\
      G2PP                & 0.56 $\pm$ 0.02 & -1.83 $\pm$ 0.14  & 3.2 $\pm$ 0.2  & 0.01 $\pm$ 0.04  & 1.97 \\
      G1PP                & 0.53 $\pm$ 0.01 & -1.92 $\pm$ 0.05  & 3.2 $\pm$ 0.2  &-0.02 $\pm$ 0.03  & 0.68 \\
      G0PP                & 0.50 $\pm$ 0.01 & -1.73 $\pm$ 0.06  & 3.2 $\pm$ 0.8  & 0.03 $\pm$ 0.04  & 0.38 \\
      \hline
      G3PPn=0                & 0.57 $\pm$ 0.02 & -2.37 $\pm$ 0.20  & 4.2 $\pm$ 0.5  & 0.04 $\pm$ 0.03  & 1.88\\ 
      G2PPn=0                & 0.57 $\pm$ 0.01 & -2.25 $\pm$ 0.12  & 4.1 $\pm$ 0.4  &-0.01 $\pm$ 0.03  & 0.95 \\
      G1PPn=0                & 0.60 $\pm$ 0.01 & -2.04 $\pm$ 0.07  & 4.9 $\pm$ 0.7  &-0.05 $\pm$ 0.02  & 0.13 \\
      G0PPn=0                & 0.52 $\pm$ 0.02 & -1.73 $\pm$ 0.06  & 3.8 $\pm$ 0.5  &-0.01 $\pm$ 0.02  & 0.03 \\
      \hline
    \end{tabular}\label{remtab}
    \medskip 
     \\
    The properties of the simulated merger remnants observed 1 Gyr 
    after the coalescence of the nuclei.
  \end{minipage}
\end{table*}

Every equal-mass gas-rich merger simulation presented here exhibits quantitatively
disturbed morphologies at some point along the merger process. However it is clear that
quantitative morphological classifications based on $G$, $M_{20}$, and $A$ are sensitive
only during the first pass and final merger stages for gas-rich equal-mass mergers, and
will miss many interacting galaxies observed between the first pass and final merger as
well as many recently merged systems. This is in qualitative agreement with the $G$, $M_{20}$, and 
$A$ values and merger stages of local ULIRGs. 
Two-thirds of the $z \sim 0.1$ ULIRG sample used to calibrate the $G-M_{20}$, $G-A$, and $C-A$ diagrams in LPM04
exhibit double or multiple nuclei, and therefore are merging systems observed at final merger stage before 
the coalescence of their nuclei or immediately at the first pass when the galaxies appear overlapping in projection 
(Figure 3).  
The $G-M_{20}$, $G-A$, and $C-A$ merger classification cuts used in LPM04 and
this work identify 93\%, 80\%, and 76\% of the double and multiple nuclei ULIRGs respectively.
The detection efficiency is significantly lower for the single nucleus ULIRGs
(46\%, 71\%, and and 54\% for $G-M_{20}$, $G-A$, and $C-A$ respectively).  This is also
in reasonable agreement with our results here, assuming that single nucleus ULIRGs are 
observed after the first pass or after the coalescence of the nuclei.  

The duration, strength, and timing of the observed morphological disturbances depend
on the merger orientation and orbital parameters, the gas properties of the initial galaxies,
and the presence of dust. When dust is included, the merger observability time-scales depend most strongly 
on the gas properties, pericentric distance, and relative orientation.   
Galaxies with high gas fractions have more star-formation along tidal arms, producing stronger asymmetries during the first pass.
Mergers with large impact parameters have long orbital decay time-scales, and exhibit disturbed
morphologies for longer. Retrograde-retrograde mergers also show disturbed morphologies
for 50-100\% longer than prograde-prograde and prograde-retrograde mergers.  
We find that the supernova feedback prescription and the total mass of the merging galaxies do not have 
a strong effect on the overall duration of morphological disturbances.
The relative orientations affect the strength of the morphological disturbances, 
with the prograde-retrograde and polar orientations showing the
strongest disturbances. The orbital parameters and gas fractions have the strongest influence on the timing of the
morphological disturbances.  Most of the high gas-fraction (Sbc) parabolic orbits show morphological disturbances at the first pass and
final merger, while the high gas-fraction highly radial orbit and large pericentric distance simulations 
have weak disturbances at the first pass and stronger disruptions at the final merger. 
The lower gas-fraction (G-series) parabolic orbit simulations experience less star-formation and morphological disturbances 
during the first pass, and hence are most likely to be detected morphologically during the final merger. 

Obscuration from dust has a very strong impact on the measured morphologies throughout the merger process
until at least 1 Gyr after the coalescence of the nuclei.  Dust extinction is highest for
the central nuclei where the star-formation rates are highest.  Because much of the central light is
masked by dust,  this results in lower $G$, $C$, and higher $M_{20}$ values during and after the merger. 
Dust lanes in the remnants can also produce higher asymmetries.   
However, the inclusion of dust does not significantly change the morphological disturbance time-scales 
during the prograde-prograde Sbc merger.  Our models may overestimate the dust content in the remnants as dust is not destroyed 
by shock-heating nor is gas removed in a post-merger `blowout' by an active galactic nucleus as predicted by other 
galaxy merger models (e.g. Hopkins et al. 2006).  However, dust destruction is expected to be most important 
after the final merger, when the star-formation rate and dust production have sufficiently declined.
Therefore, while the dust and gas content of our merger remnants may be overestimated, the morphologies
and time-scales calculated during the merger are unlikely to be affected by the destruction/removal of
dust at late stages. 

The observability time-scales clearly depend on the method used to select merger candidates.
The time-scale during which a merging system is a close pair at a particular projected separation
is not the same as the time-scale during which the system shows high asymmetries or high $G-M_{20}$ values. 
The gas-rich Sbc mergers have 2-4 times longer $T(G-A)$ and $T(A)$ 
than $T(G-M_{20})$ and $T( 10< R_{proj} < 30 h^{-1}$ kpc).   The lower gas-fraction G-series simulations, on the
other hand,  have similar time-scales for $G-M_{20}$, $G-A$, and $A$ disturbances. Although $G$ and  $M_{20}$ are the most 
affected by dust, the $G-M_{20}$ time-scale is the least affected by the merger parameters with 
typical time-scales $\sim 0.2-0.3$ Gyr for the G-series simulations and
$\sim 0.3-0.6$ Gyr for the Sbc simulations.
The typical $G-A$ time-scales are $\sim 0.3-0.4$ Gyr for the G-series simulations and $\sim 0.8-1.1$ Gyr for the Sbc simulations.  
The typical $A$ time-scales are $\sim 0.2-0.3$ Gyr for the G-series simulations and $\sim 0.7-1.1$ Gyr for the Sbc simulations.  
Thus the $G-M_{20}$ time-scales have $\sim$ 0.4 Gyr dispersion,  while the $G-A$ and $A$ time-scales have $\sim$ 0.8 Gyr dispersion.    
While more mergers may be identified using asymmetry given the longer asymmetry time-scales,
the merger rate calculated using $G-M_{20}$ mergers will be less uncertain given the better consistency of the $G-M_{20}$ time-scales. 

The close pair time-scales depend on the orbital decay times,  
with the smaller projected separations showing the greatest fractional variability. 
For $5 < R_{proj} < 20 h^{-1}$ kpc, the observability time-scales are $\sim 0.4-0.6$ Gyr for the G-series simulations
and $\sim 0.1-0.3$ Gyr for Sbc.   These are slightly longer at $10< R_{proj} < 30 h^{-1}$ kpc, with time-scales  
$\sim 0.5-0.7$ Gyr for the G-series simulations and $\sim 0.2-0.5$ Gyr for Sbc.   At larger projected radii $<$ 50, 100 $h^{-1}$ kpc, 
the highly radial orbit and the large pericentric distance parabolic orbit have significantly shorter ($\sim$ 0.3, 0.8 Gyr) 
and longer time-scales ($\sim$ 1.4, 3.0 Gyr), respectively.   The typical $10< R_{proj} < 50 h^{-1}$ kpc 
time-scales are $\sim 0.7-1.2$ Gyr for the G3/G2 simulations and $\sim 0.9-1.1$ Gyr for the Sbc simulations. 
The typical $10 < R_{proj} < 100 h^{-1}$ kpc time-scales are $\sim$ 1.9 Gyr for the G3 simulations 
and $\sim 1.1-1.4$ Gyr for the Sbc simulations.   Therefore the observability time-scales for close pairs 
vary by a factor of 4-6 at small projected separations and by a factor of 2-3 at larger separations.  
It is important to keep in mind that these are the observability time-scales for truly merging pairs  ($T_{pair}$ in Eqn. 16) 
and does not include the contamination correction for non-merging pairs observed in projection ( $p(merg)$ in Eqn. 16). 
Close pairs with large projected separations are more likely to be contaminated by non-merging galaxies, 
so the optimal separation distance is likely to be at intermediate separations between 30-50 $h^{-1}$ kpc. 

Unlike our simulated mergers where we know the merger parameters and initial galaxy properties a priori, 
it is generally impossible to recover these for each galaxy merger observed in large surveys of  the distant universe. 
Ideally, one would like to determine an effective observability time-scale for each method of identifying mergers 
in order to convert the number density of observed mergers observed into a galaxy merger rate.
This effective time-scale should be weighted by the distribution of initial galaxy properties, mass ratios, and orbital
parameters predicted for galaxy mergers by cosmological simulations.    Our work here is a first step towards determining
the mean observability time-scale and has concentrated on the systems most likely to be affected by dusty starbursts, i.e. gas-rich
equal mass mergers of discs with small bulges. While this is significant improvement over previous estimates of the morphological
disturbance time-scales, it is not sufficient to convert the observed fraction of morphologically disturbed and paired galaxies
into a galaxy merger rate. Our next paper will explore the merger observability time-scales
of non-equal mass mergers needed to estimate the effective observability time-scale for a realistic population of 
mergers and calculate the galaxy merger rate. 

Galaxy mergers are often assumed to be associated with vigorous starbursts, such as observed 
for local ultra-luminous infrared galaxies (e.g. Sanders \& Mirabel 1996).
Our results here suggest that the timing of morphological disturbances can be offset from the
peak in star-formation rate, especially if  the stiff $n=2$ supernova feedback prescription is correct.  
In general, the maximum morphological disturbances occur  before the
peaks in the star-formation rate.   While asymmetries experience a sharp peak at the first pass lasting 100-200 Myr, 
the star-formation rate of the system remains enhanced above the initial rate for significantly longer after the first pass 
for both supernova feedback models.  Asymmetries also peak sharply at the final merger, while the
star-formation peaks after the galaxies appear as single object when the nuclei coalesce.  
 Therefore, the objects with highest star-formation rates may not always have the highest asymmetries. 
This is particularly true for the stiff $n=2$ supernova feedback models, which  experience a long burst at the final merger
($\sim$ 500 Myr) lasting longer than the high asymmetries 
($\sim$ 200 Myr). During both first pass and final merger,  the majority of the enhanced 
star formation occurs in the nucleus which makes the quantitative morphologies appear more concentrated but 
not necessarily disturbed. If we ignore the effect of dust on the morphologies, $G$
correlates directly with the star-formation rate. Because dust lowers the measured $G$ value, 
the dusty Sbc final mergers do not show high $G-M_{20}$, while the less-obscured G-series final mergers do 
(Figs. \ref{merg_orient}, \ref{merg_mass}).  

Roughly 75\% of the strongest starbursts in the local universe, ULIRGs, have quantitatively disturbed morphologies, and
two-thirds show multiple nuclei.  Therefore the correlation between disturbed morphologies 
and peak star-formation rates appears to be better than what is implied by our models. 
We note that only one of our simulations (the gas-rich,  radial orbit, isothermal feedback simulation SbcRn=0) reaches a 
star-formation rate at the final merger that is 
comparable to ULIRGs. ULIRGs may have higher gas fractions than our models, and 
often host active nuclei that could destroy or sweep out dust in the central regions 
during the final merger, both of which would tighten the correlation between disturbed morphologies and high star-formation rates. 
Finally, the star-formation rates of some ULIRGs may be overestimated, as 
some of the infrared luminosity may be from dust-heating by an AGN rather than star-formation. 
Any correlation between high star-formation rates and disturbed morphologies will depend on the star-formation indicator
used to calculate the star-formation rate.  Observed H$\alpha$ luminosities and equivalent widths are expected to be highest
during the first pass and initial starbursts before the star-forming regions have been enshrouded in dust, 
while infrared luminosities will peak during the final merger after sufficient amounts of metals and dust have been produced 
(Jonsson et al. 2006).  

\begin{figure} 
\includegraphics[height=225mm]{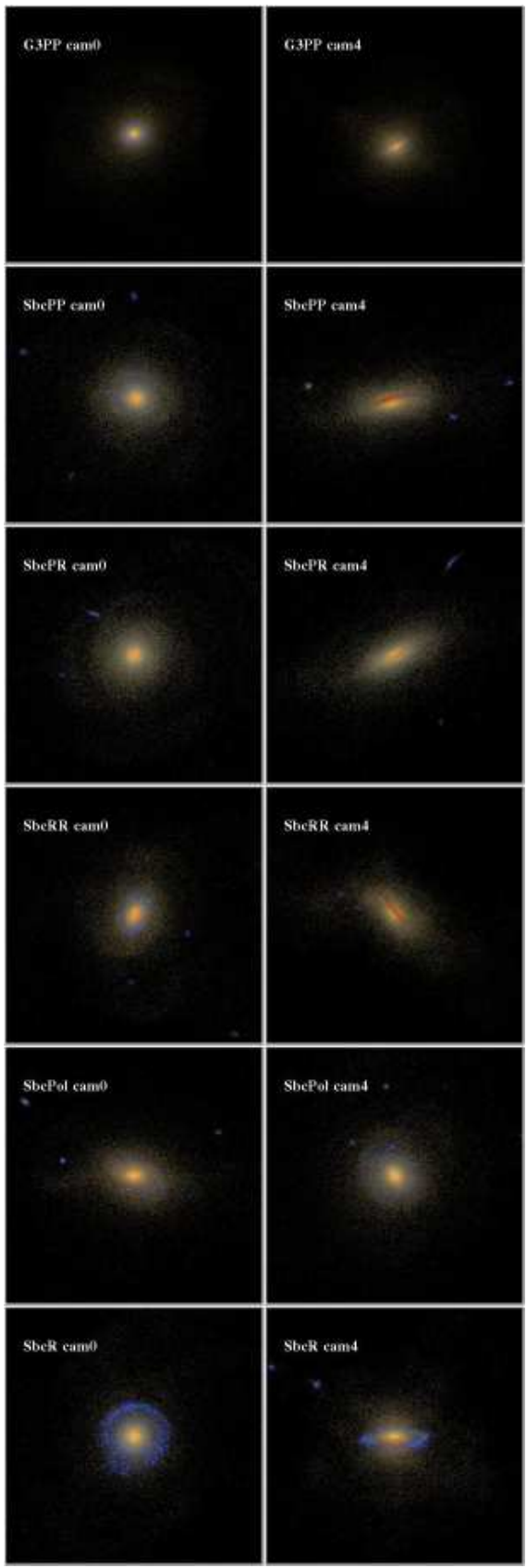}
\caption{Left: G3 and Sbc remnants viewed from cam0. Right: G3 and Sbc remnant viewed from cam4.  All of the remnants
show a low-mass dusty star-forming disc as well as a large bulge component.}
\end{figure}

The vast majority of our equal-mass gas-rich merger remnants are decidedly disc-like and dusty, even those mergers which
started with relatively low gas fractions. Previous studies of these and other equal-mass merger simulations
have shown that the mass distribution of star particles follow $r^{1/4}$ laws with steep central cusps 
consistent with or more concentrated than the light profiles of elliptical galaxies (e.g. 
Cox et al. 2006, 2008; Naab et al. 2006; Bournaud, Jog, \& Combes 2005)
and lie on the fundamental plane (Robertson et al. 2006b).  Only simulations with gas fractions $>$ 50\%
(Springel \& Hernquist 2005; Robertson et al. 2006a) or mass ratios less than 1:3 (Naab et al. 2006, 
Bournard et al. 2005) have been found to have merger remnants with massive disc components. 
The masses of our merger remnants are dominated by the bulge component; 
our  merger remnants appear disc-like and less concentrated than elliptical galaxies because we examine the $g$-band light
profiles which are strongly affected by both dust extinction in the central regions and a bright but low-mass disc
of young stars.   Only models with less dusty remnants (as produced by $n=0$ supernova feedback) 
or models that ignore dust in the remnant
produce remnants with high enough $G$, $C$, and low enough $M_{20}$ values to be called spheroids.  However,
even these spheroidal remnants are forming stars at rates $>$ 1 M$_{\odot}$ yr$^{-1}$, and would not be classified
as red E/S0 or post-starburst E+A galaxies. 
If these simulations accurately represent the end stages of the merger process, 
a number of `green' Sb galaxies may be merger remnants (e.g.  Hammer et al. 2005).   
However, our models may overestimate the extinction  and star-formation 
during the post-merger stages.  Destroying dust
by shock-heating would not be sufficient to produce true red and dead spheroids, and 
additional physics such as feedback from an active galactic nucleus may be needed to clear out the gas and kill star formation
(see also Khalatyan et al. 2008). 
The remnants forming the fewest stars and with the least dust are those produced by the lowest mass systems.  
But the low-mass remnants are the least centrally
concentrated of all the simulated remnants. 
Like their higher mass counterparts, the low-mass remnants have more gas ($\sim$ 10$^8$
M$_{\odot}$) and star formation than typical dwarf ellipticals. 

\section{SUMMARY}
We present a morphological analysis of a large suite of GADGET N-Body/SPH equal-mass gas-rich disc galaxy merger
simulations which have been processed through the Monte-Carlo radiative transfer code SUNRISE.  With the resulting images, 
we have examined the dependence of quantitative morphology and projected separation in the SDSS $g$-band on 
merger stage, dust, viewing angle, merger orientation and orbital parameters, gas properties, supernova feedback prescription,
and total mass.  We have determined the time-scales for quantitative morphological disturbances in the Gini coefficient,
$M_{20}$, $C$, and $A$, and the time-scale during which close pairs lie at projected separations $R_{proj} <$ 20, 30, 50, 
and 100 $h^{-1}$ kpc.   We also examine the merger remnant morphologies and star-formation rates. 

$\bullet$ All of the equal-mass gas-rich merger simulations experience quantitatively disturbed morphologies in
$G-M_{20}$, $G-A$, and $A$ at the first pass and/or the final merger.  This is in good agreement with the morphologies and
merger stages of the local ULIRG sample used to empirically calibrate these quantities.  However, merging galaxies observed
between the first pass and final merger or after the coalescence of their nuclei may not show disturbed $G-M_{20}$ and asymmetries. 

$\bullet$ The time-scale during which an equal-mass gas-rich merger may be identified is strongly dependent
on the method used to find the merger.  The $G-M_{20}$ time-scales are the shortest of the morphological methods, but 
have the least dependence on the merger parameters with $T(G-M_{20}) \sim 0.2-0.6$ Gyr.  The asymmetry time-scales
vary by a factor of 3-4 between $\sim 0.2-1.1$ Gyr, and the $G-A$ time-scales are the longest, with $T(G-A) \sim 0.3-1.2$ Gyr.
The close pair time-scales vary by factor of 2$-$6 with the orbital parameters,  depending on the projected separations adopted. 
At $5 < R_{proj} < 20 h^{-1}$ kpc, the observability time-scales are $\sim 0.1-0.6$ Gyr.  At $10 < R_{proj} < 30 h^{-1}$ kpc, 
the observability time-scales are $\sim 0.2-0.7$ Gyr.   At $10 < R_{proj} < 50 h^{-1}$ kpc, 
the typical observability time-scales are $\sim 0.7-1.2$ Gyr.   At $10 < R_{proj} < 100 h^{-1}$ kpc, 
the typical observability time-scales are $\sim 1.1-1.9$ Gyr.

$\bullet$ The presence of dust has strong impact on the quantitative morphological measurements, 
lowering $G$ and $C$, raising $M_{20}$ throughout the merger, and raising $A$ during the post-merger and remnant
stages. 

$\bullet$ When dust is included, the time-scales for morphological disturbances are most sensitive to the gas fraction of the 
merging galaxies, their pericentric distance, and relative orientation.   The supernova feedback prescription 
and the total mass of system do not significantly change the morphological time-scales.
The relative orientations also affect the strength of the morphological disturbances, with prograde-retrograde and polar
orientation showing the highest asymmetries.    The timing of the disturbances also depends on orbital parameters and
gas fractions, with low gas fractions, large pericentric distances, and highly radial orbits showing strong disturbances primarily
during the final merger.

$\bullet$ The timing of morphological disturbances is generally offset from the peak in star-formation rates, with strong
morphological disturbances occurring before bursts of merger-induced star-formation and for shorter durations. Hence, not
all merger-induced starbursts will exhibit morphological disturbances and vice versa. The mode of supernova feedback and
dust production also play important roles in the correlation between morphological disturbances and observed star-formation indicators.

$\bullet$ The majority of simulated merger 
remnants observed $\ge$ 1 Gyr after the coalescence of their nuclei 
appear disc-like and dusty in $g$-band light and are consistent with early-type spiral morphologies 
and star-formation rates. Decreased dust extinction would make most remnants appear more spheroidal, 
but would not affect the remnants' high star-formation rates (typically $>$ 1 M$_{\odot}$ yr$^{-1}$).  A major gas-rich merger 
without AGN feedback does not, by itself, produce a red and dead spheroidal galaxy. 

We  would like to thank our referee, Chris Conselice, for his helpful comments. 
JML acknowledges support from the NOAO Leo Goldberg Fellowship, NASA grants NAG5-11513 and HST-AR-9998, and 
would like to thank P. Madau for support during of this project. 
PJ was supported by programs HST-AR-10678 and HST-AR-10958, provided by NASA through grants from the Space Telescope Science
Institute, which is operated by the Association of Universities for Research in Astronomy, Incorporated, 
under NASA contract NAS5-26555, and  by the Spitzer Space Telescope Theoretical Research Program, through a contract 
issued by the Jet Propulsion Laboratory, California Institute of Technology under a contract with NASA.
TJC was supported by a grant from the W.M. Keck Foundation. 

This research used computational resources of the
NASA Advanced Supercomputing Division (NAS) and the National Energy
Research Scientific Computing Center (NERSC), which is supported by
the Office of Science of the U.S. Department of Energy.


\begin{thebibliography}{}
\bibitem[]{} Abraham, R. G., Valdes, F., Yee, H.K.C., \& van den Bergh, S., 1994, ApJ, 432, 75
\bibitem[]{} Abraham, R., van den Bergh, S., \& Nair, P. 2003, ApJ, 588, 218
\bibitem[]{} Abraham, R. et al. 2007, ApJ, 669, 184
\bibitem[]{} Barton, E.J., Geller, M.J., \& Kenyon, S.J. 2000, ApJ, 530, 660
\bibitem[]{} Bell, E.F. \& de Jong, R.S. 2001, ApJ, 550, 212
\bibitem[]{} Bell, E.F., McIntosh, D.H., Katz, N., Weinberg, M.D. 2003, ApJL, 585, L117
\bibitem[]{}Bell, E.F., et al. 2004, ApJ, 608, 752
\bibitem[]{}Bell, E.F., Phelps, S., Somerville, R., Wolf, C., Borch, A., \& Meisenheimer, K. 2006a, ApJ, 652, 270
\bibitem[]{}Bell, E.F. et al. 2006b, ApJ, 640, 241
\bibitem[]{}Berrier, J.C., Bullock, J.S., Barton, E.J., Guenther, H.D., Zentner, A., \& Wechsler, R. 2006, ApJ, 652, 56
\bibitem[]{}Bershady, M., Jangren, A., \& Conselice, C. 2000, AJ, 119, 2645
\bibitem[]{}Bertin, E. \& Arnouts, B. 1996,  A\&AS, 117, 393
\bibitem[]{}Borne, K. D., Bushouse, H., Lucas, R. A., \& Colina, L. 2000, ApJ, 529, 77L
\bibitem[]{}Boylan-Kolchin, M., Ma, C.-P., \& Quataert, E. 2008, MNRAS, 383, 93
\bibitem[]{}Boylan-Kolchin, M., Ma, C.-P., \& Quataert, E. 2005, MNRAS, 362, 184
\bibitem[]{}Bournaud, F., Jog, C.J., \& Combes, F. 2005, A\&A, 437, 69
\bibitem[]{}Brinchmann, J. et al. 1998, ApJ, 499, 112
\bibitem[]{}Broeils, A.H. \& van Woerden, H. 1994, A\&AS, 107, 129
\bibitem[]{}Brown, M. et  al. 2007, ApJ, 654, 858
\bibitem[]{}Bundy, K., Ellis, R.S., \& Conselice, C. 2005, ApJ, 625, 621 
\bibitem[]{}Conselice, C. Bershady, M.A., Jansen, A. 2000, ApJ, 529, 886 
\bibitem[]{}Conselice, C., Bershady, M.A., Dickinson, M., \& Papovich, C. 2003, AJ, 126, 1183
\bibitem[]{}Conselice, C. 2003, ApJS, 147, 1
\bibitem[]{}Conselice, C., Blackburne,  J., \& Papovich, C. 2005, ApJ, 620, 564
\bibitem[]{}Conselice, C. 2006, ApJ, 638, 686
\bibitem[]{}Cox, T.J., Primack, J.R., Jonsson, P., \& Somerville, R.S. 2004, ApJL, 607, L87
\bibitem[]{}Cox, T.J., Jonsson, P., Primack, J.R., Somerville, R.S. 2006, MNRAS, 373, 1013
\bibitem[]{}Cox, T.J., Jonsson, P., Somerville, R.S., Primack, J.R., \& Dekel, A. 2008, MNRAS, 384, 386
\bibitem[]{}Dasyra, K.M., et al. 2006, ApJ, 638, 745
\bibitem[]{}Dekel, A. \& Birnboim, Y. 2006, MNRAS, 368, 2
\bibitem[]{}de Jong, R.S. 1996, A\&A, 313, 45
\bibitem[]{}de Propris, R. et al. 2005, AJ, 130, 1516
\bibitem[]{}Fanning, D. 2002, {\it Coyote's Guide to IDL Programming}
\bibitem[]{}Faber, S.M. et al. 2007, ApJ, 665, 265
\bibitem[]{}Glasser, G.J. 1962, Amer. Stat. Assoc. 57, 648, 654
\bibitem[]{}Guo, Q. \& White, S. 2008, MNRAS, 384, 2
\bibitem[]{}Hammer, F., Flores, H., Elbaz, D., Zheng, X.Z., Liang, Y.C., \& Cesarsky, C. 2005, A\&A, 430, 115
\bibitem[]{}Hogg, D. Masjedi, M., Berlind, A., Blanton, M., Quintero, A., \& Brinkmann, J. 2006, ApJ, 650, 763
\bibitem[]{}Hopkins, P., Herniquist, L, Cox, T.J., Di Matteo, T., Robertson, B., \& Springel, V. 2006, ApJS, 163, 1
\bibitem[]{}Hopkins, P., Bundy, K., Hernquist, L., \& Ellis, R. 2007, ApJ, 659, 976
\bibitem[]{}Iono, D. Yun, M.S. \&  Mihos, J.C. 2004, ApJ, 616, 199
\bibitem[]{}Jiang, C.Y., Jing, Y.P., Faltenbacher, A., Lin, W.P., \& Li, C., 2008, ApJ, 675, 1095
\bibitem[]{}Jogee, S. et al. 2008, submitted to ApJ
\bibitem[]{}Jonsson, P. 2006, MNRAS, 372, 2
\bibitem[]{}Jonsson, P., Cox, T.J., Primack, J., \& Somerville R. 2006, ApJ, 637, 255
\bibitem[]{}Katz, N., Weinberg, D.H., Hernquist, L. 1996, ApJS, 105, 19
\bibitem[]{}Kampczyk, P. et al. 2007, ApJS, 172, 329
\bibitem[]{}Kartaltepe, J.S. et al. 2007, ApJS, 172, 320
\bibitem[]{}Kauffmann, G., White, S.D.M., \& Guiderdoni, B. 1993, MNRAS, 264, 201
\bibitem[]{}Kennicutt, R.C. 1998, ApJ, 498, 541
\bibitem[]{}Kere\v{s}, D., Katz, N., Weinberg, D.H., \& Dave, R. 2005, MNRAS, 363, 2
\bibitem[]{}Khalatyan, A., Cattaneo,  A., Schramm, M., Gottlober, S., Steinmetz, M., \& Wisotzki, L. 2008, MNRAS, 387, 13
\bibitem[]{}Kitzbichler, M. \& White, S. 2008, submitted to MNRAS, arXiv0804.1965
\bibitem[]{}Law, D. et al. 2007,  ApJ, 656, 1L
\bibitem[]{}Leitherer, C. et al. 1999, ApJS, 123, 3
\bibitem[]{}Li, C., Kauffmann, G., Heckman, T., Jing, Y.P., \& White, S.D.M. 2008, MNRAS,  385, 1903
\bibitem[]{}Lin, L. et al. 2004, ApJ, 617, L9
\bibitem[]{}Lin, L. et al. 2008, ApJ, 681, 232
\bibitem[]{}Lorenz, M.O. 1905, Amer. Stat Assoc., 9, 209
\bibitem[]{}Lotz, J.M., Primack, J., \& Madau, P. 2004, AJ, 613, 262 (LPM04)
\bibitem[]{}Lotz, J.M. et al 2008, ApJ, 672, 177
\bibitem[]{}Masjedi, M., Hogg, D.W., \& Blanton, M.R. 2008, 679, 260
\bibitem[]{}Masjedi, M. et al. 2006, ApJ, 644, 54
\bibitem[]{}Mihos, J.C., \& Hernquist, L. 1996, 464, 641
\bibitem[]{}Moore, B., Lake, G., \& Katz, N. 1998, ApJ, 495, 139
\bibitem[]{}Naab, T., Jesseit, R., \& Burkert, A., 2006, MNRAS, 372, 839
\bibitem[]{}Naab, T., Khochfar, S., \& Burkert, A., 2006, ApJ, 636, 81
\bibitem[]{}Patton, D et al. 2000, ApJ, 536, 153
\bibitem[]{}Patton, D.R. et al. 2002, ApJ, 565, 208 
\bibitem[]{}Petrosian, V. 1976, ApJ, 209, L1
\bibitem[]{}Ravindranath, S. et al. 2006, ApJ, 652, 963
\bibitem[]{}Renzini, A. 2007, in {\it At the Edge of the Universe: Latest Results from Deepest Astronomical Surveys}, ASP Conf. Ser., ed. J. Afonso, H.C. Ferguson, B. Mobasher, \& R. Norris, p. 309
\bibitem[]{}Rocha, M.,  Jonsson, P.,  Primack, J.R., \& Cox, T.J. 2008, MNRAS, 383, 1281
\bibitem[]{}Roberts, M.S. \& Haynes, M.P. 1994, ARA\&A, 32, 115
\bibitem[]{}Robertson, B., Bullock, J., Cox, T.J., di Matteo, T., Hernquist, L., Springel, V., \& Yoshida, N. 2006, ApJ, 645, 986 
\bibitem[]{}Robertson, B., Cox, T.J., Hernquist, L., Franx, M., Hopkins, P., Martini, P., \& Springel, V., ApJ, 641, 21
\bibitem[]{}Ryan, R.E., Cohen, S.H., Windhorst, R.A., \& Silk, J. 2008, ApJ, in press
\bibitem[]{}Sanders, D.B. \& Mirabel, I.F. 1996, ARA\&A, 34, 749
\bibitem[]{}Scarlata, C. et al. 2007, ApJS, 172, 406 
\bibitem[]{}Shen, S. et al. 2003, MNRAS, 343, 978
\bibitem[]{}Shi, Y. 2008, in preparation
\bibitem[]{}Somerville, R.S. Primack, J.R.\& Faber, S.M. 2001, MNRAS, 320, 504
\bibitem[]{}Springel, V., Yoshida, N., \& White, S.D.M., 2001, New Astronomy, 6, 79
\bibitem[]{}Springel, V. \& Hernquist, L. 2002, MNRAS, 333, 649
\bibitem[]{}Springel, V. \& Hernquist, L. 2003, MNRAS, 339, 289
\bibitem[]{}Springel, V. \& Hernquist, L. 2005, ApJ, 622, L9 
\bibitem[]{}Stewart, K., Bullock, J., Wechsler, R., Maller, A.H., \& Zentner, A. 2008, ApJ, 683, 597
\bibitem[]{} Takamiya, M. 1999, ApJS, 122, 109
\bibitem[]{}Toomre, A. 1977, in {\it Evolution of Galaxies and Stellar Populations}, eds. B.M. Tinsely \& R. Larson, (New Haven: Yale University Obs.), 401
\end{thebibliography}
\end{document}